\documentclass[a4]{amsart}

\usepackage{rotating}

\usepackage[noend]{algorithmic}
\usepackage{algorithm}
\usepackage{float}
\floatname{algorithm}{Algorithm}
\floatname{dataStructure}{Data structure}

\newcommand{\hadgesh}[1]{{\emph{#1}}}

\input xy
\xyoption{all}

\usepackage{amssymb,amsmath}
\newtheorem{Thm}{Theorem}
\newtheorem{Ex}[Thm]{Example}

\newcommand{\cals}{\mathcal{S}}
\newcommand{\F}{{\mathbb F}}
\newcommand{\rk}{\operatorname{rk}\nolimits}
\newcommand{\nll}{\operatorname{null}\nolimits}
\newcommand{\Ker}{\operatorname{Ker}\nolimits}
\newcommand{\calg}{\mathcal{G}}
\newcommand{\calh}{\mathcal{H}}
\newcommand{\argmax}{\operatornamewithlimits{argmax}}

\begin{document}

\title[Topological analysis of neural microcircuits]{Topological analysis of the connectome of digital reconstructions of neural microcircuits}

\author[D\l otko et al.]{Pawe\l\ D\l otko${}^{*, 1}$}
\thanks{${}^{*}$co-first author and corresponding author}
\thanks{(1) Partial support  provided by the Advanced Grant of the European Research Council GUDHI (Geometric Understanding in Higher Dimensions)}
\author[]{Kathryn Hess${}^{*}$}
\author[]{Ran Levi${}^{*}$}
\author[]{Max Nolte${}^{*}$}
\author[]{Michael Reimann}
\author[]{Martina Scolamiero}
\author[]{Katharine Turner}
\author[]{Eilif Muller}
\author[]{Henry Markram}

\address{Geometrica, Inria, Saclay, France}
\address{Laboratory for Topology and Neuroscience, \'Ecole Polytechnique F\'ed\'erale  de Lausanne, Lausanne, Switzerland}
\address{Institute of Mathematics, University of Aberdeen, Aberdeen, UK}
\address{Blue Brain Project, \'Ecole Polytechnique F\'ed\'erale de Lausanne, Lausanne, Switzerland}
\address{Blue Brain Project, \'Ecole Polytechnique F\'ed\'erale de Lausanne, Lausanne, Switzerland}
\address{Laboratory for Topology and Neuroscience, \'Ecole Polytechnique F\'ed\'erale  de Lausanne, Lausanne, Switzerland}
\address{Laboratory for Topology and Neuroscience, \'Ecole Polytechnique F\'ed\'erale  de Lausanne, Lausanne, Switzerland}
\address{Blue Brain Project, \'Ecole Polytechnique F\'ed\'erale de Lausanne, Lausanne, Switzerland}
\address{Blue Brain Project, \'Ecole Polytechnique F\'ed\'erale de Lausanne, Lausanne, Switzerland}

\begin{abstract}
{A recent publication provides the network graph for a neocortical microcircuit comprising 8 million connections between 31,000 neurons \cite{BBP-ref}. Since traditional graph-theoretical methods may not be sufficient to understand the immense complexity of such a biological network, we explored whether methods from algebraic topology could provide a new perspective on its structural and functional organization. Structural topological analysis revealed that directed graphs representing connectivity among neurons in the microcircuit deviated significantly from different varieties of randomized graph. In particular, the directed graphs contained in the order of $10^7$ simplices Ð groups of neurons with all-to-all directed connectivity. Some of these simplices contained up to 8 neurons, making them the most extreme neuronal clustering motif ever reported. Functional topological analysis of simulated neuronal activity in the microcircuit revealed novel spatio-temporal metrics that provide an effective classification of functional responses to qualitatively different stimuli. This study represents the first algebraic topological analysis of structural connectomics and connectomics-based spatio-temporal activity in a biologically realistic neural microcircuit. The methods used in the study show promise for more general applications in network science.}
\end{abstract}


\maketitle
\begin{figure*}
\centering
\includegraphics[scale = 0.35,angle=0.00, clip=true]{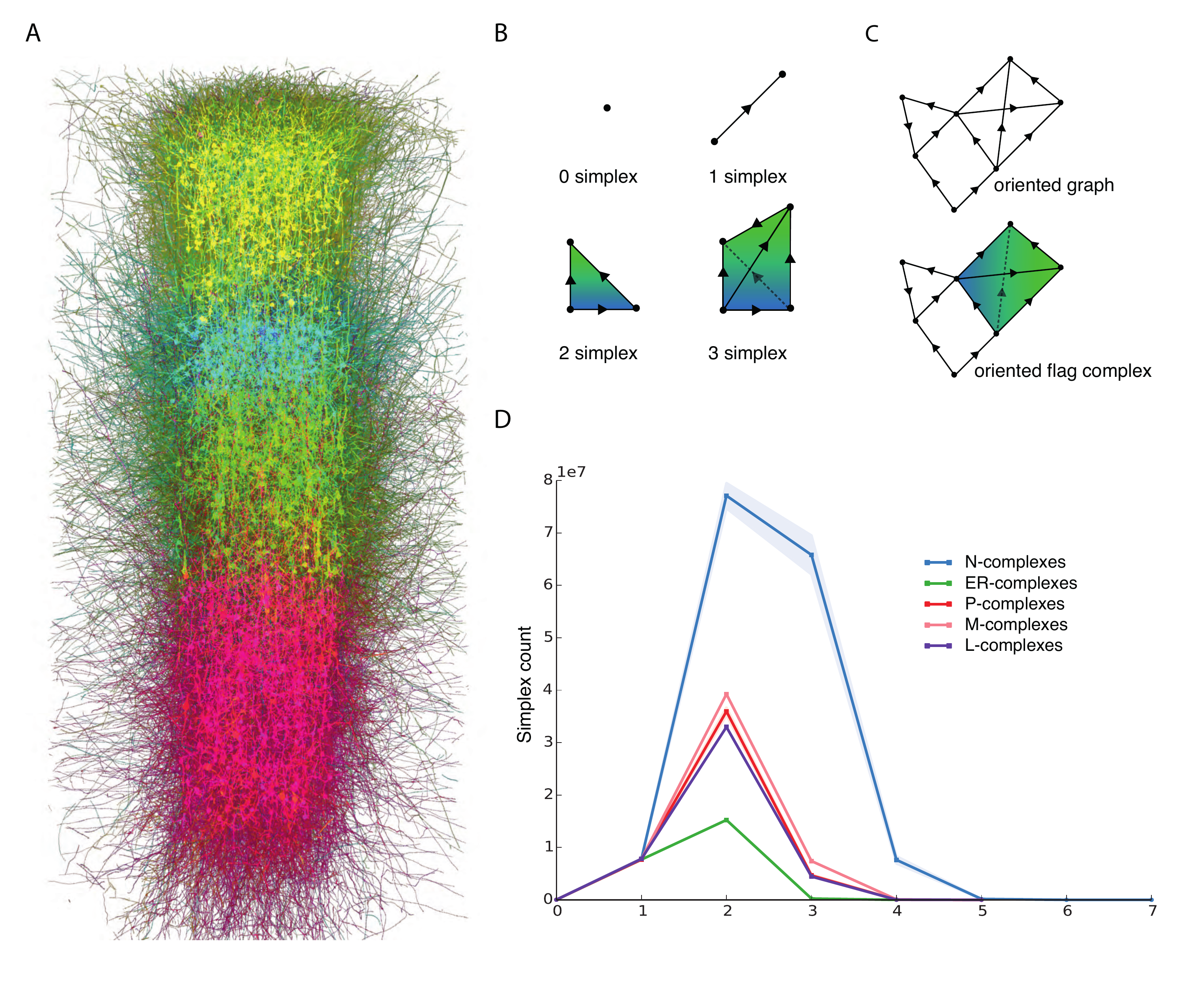}
\caption{\emph{(A)} A sparse visualization of the microcircuit (soma and dendrites only). Morphological types are color-coded, with m-types in the same layer having similar colors. \emph{(B)} Examples of simplices in dimensions 0 through 3. \emph{(C)} An example of a directed graph and its associated flag complex, in which there is one $n$-simplex for every directed $(n+1)$-clique in the graph. \emph{(D)} A graph depicting the average  number of simplices in each dimension for the flag complexes associated to the reconstructed microcircuit (N-complexes) and the four types of random graphs considered, each with the same number of vertices as the reconstructed microcircuit, where shading indicates standard deviation, which was very small for all except the N-complexes.}
\label{fig:figure1}
\end{figure*}

The Blue Brain Project (BBP) has recently generated the first draft digital reconstruction and simulation of a microcircuit of neurons in the neocortex of a two-week-old rat (Figure \ref{fig:figure1}A) \cite{BBP-ref}.  This reconstruction is made available through the Neocortical Microcircuit Portal (https://bbpnmc.epfl.ch) \cite{nmc}. Based on sparse anatomical and physiological data for neurons and synapses and on a variety of biologically motivated organizing principles, the complete connectivity between neurons belonging to a neocortical microcircuit was digitally reconstructed -- a Òmicro-connectomeÓ. The structural properties of the reconstruction have been extensively validated against independent data, and simulations of the reconstruction reproduced multiple in vitro and in vivo experiments without adjusting any parameter, further validating its biological accuracy.

In this article we apply methods from topology to the analysis of 42 variants of the digital reconstruction, grouped in six sets of seven microciruits each. The first five sets of microcircuits take into account biological variability in layer heights, proportions of cell types, and cell densities from five individual rats, while the sixth set is based on the average reconstruction across the five individuals. To form each set of microcircuits, seven statistically varying instantiations of the microcircuit were reconstructed \cite{reimann}. The 42 microcircuits are therefore all distinct, though the degree of resemblance within each set is higher than that between sets. The structural connectivity of each reconstructed microcircuit can be represented as a directed graph with approximately $3\times 10^{4}$ vertices and $8\times 10^{6}$  edges, while its functional connectivity can be represented as a time series of subgraphs formed by functionally effective connections. 

\begin{figure*}
\centering
\includegraphics[scale = 0.35,angle=0.00, clip=true]{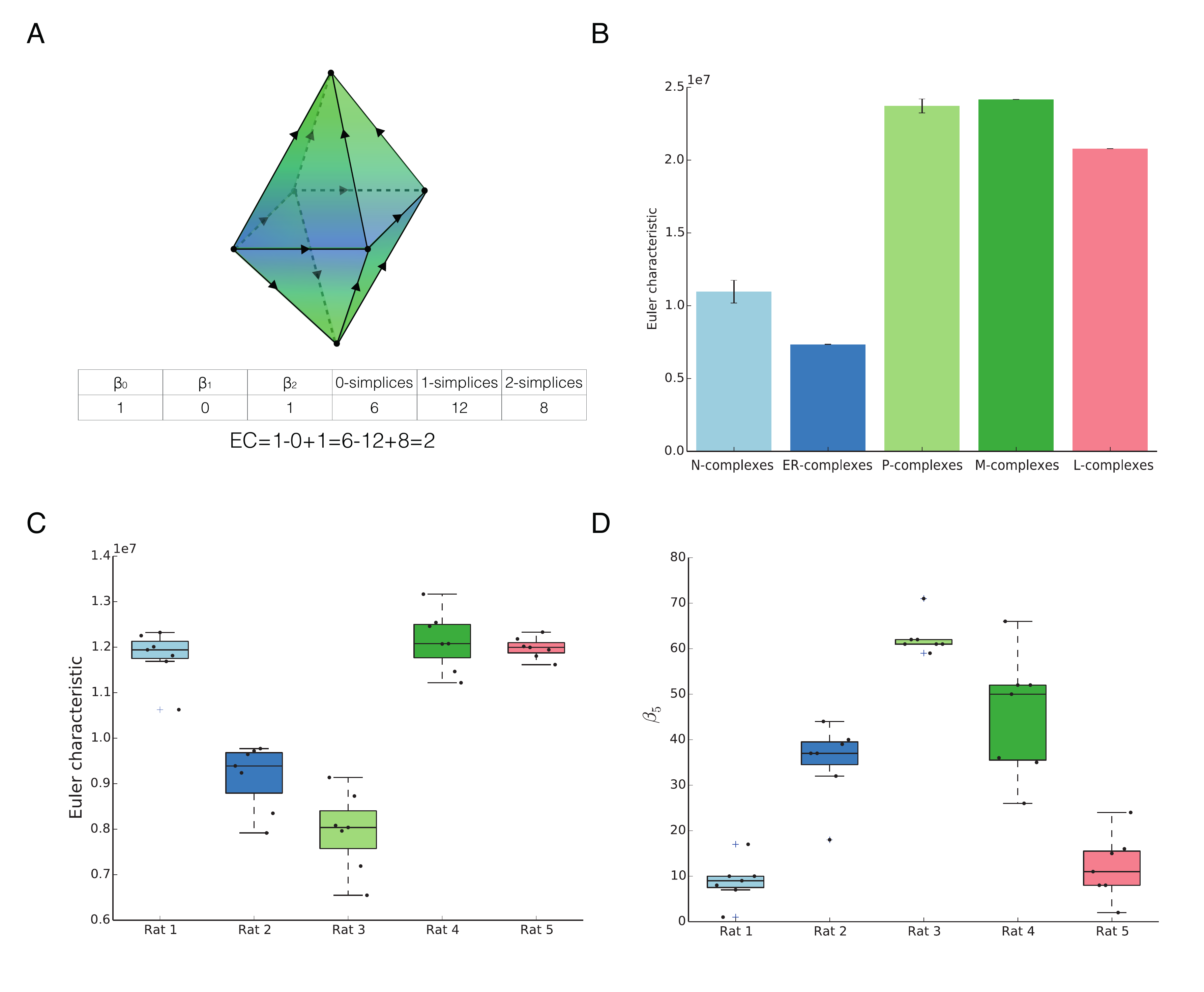}
\caption{\emph{(A)} An oriented simplicial complex consisting of eight 2-simplices glued together along their 1-dimensional faces, together with a table of its Betti numbers and numbers of simplices in dimensions 0,1, and 2 and a computation illustrating that the Euler characteristic can be computed as the alternating sum of the Betti numbers or the simplex counts. \emph{(B)}  Graph depicting the average Euler characteristic of the reconstructed microcircuit (N-complexes) and of each of the types of random graph considered, where the whisker indicates standard deviation, which was very small, except for N-complexes and P-complexes.  \emph{(C)}  Box-and-whisker plots depicting the Euler characteristics of 35 reconstructed microcircuits, seven for each individual rat.  \emph{(D)} Box-and-whisker plots depicting the 5th Betti number of 35 reconstructed microcircuits, seven for each individual rat.
}
\label{fig:figure2}
\end{figure*}

Our topological analysis of the detailed structural and functional connectivity of these 42 neural microcircuits led to a number of surprising observations.   Firstly, we found that the distribution of directed cliques (directed all-to-all connected subsets) of neurons by size is highly significantly different from both that in Erd\H os-R\'enyi random graphs with the same number of vertices and the same average connection probability and that in more sophisticated random graphs, constructed either by taking into account distance-dependent probabilities varying within and between cortical layers or morphological types of neurons, or according to Peters' Rule \cite{Peters1}, \cite{Peters2} (Figure \ref{fig:figure1}D). In particular, we found that directed cliques of up to eight neurons are highly prominent motifs in the reconstructed microcircuits: the average microcircuit incorporates approximately $10^{8}$ $3$-cliques and $4$-cliques, approximately $10^{7}$ $5$-cliques, approximately $10^{5}$ $6$-cliques, and approximately $10^{3}$ $7$-cliques.  Taking the alternating sum of the numbers of directed cliques of various sizes, we computed the \hadgesh{Euler characteristic (EC)} \cite{hatcher} of the 42 reconstructed microcircuits, obtaining in each case a value on the order of $10^{7}$, indicating a preponderence of directed cliques consisting of an odd number of neurons (Figure \ref{fig:figure2}).

Another topological metric that we considered in this analysis are the \emph{Betti numbers} (SI, Supplementary Text, ST1.3) associated to a graph via its \emph{directed flag complex} (Figure \ref{fig:figure1}C).   These are a sequence of natural numbers $\beta_{0}, \beta_{1}, \beta_{2},...$ that measure the higher-order organizational complexity of the network, detecting ``cyclic'' chains of intersecting directed cliques.  For each graph considered here we determined its \emph{homological dimension}, i.e., the maximum $n$ such that $\beta_{n}\not=0$.  We showed that the reconstructed microcircuits have homological dimension 5 (Figure \ref{fig:figure2}D), whereas the random graphs considered have homological dimension at most 4, strongly indicating that the microcircuits possess a higher degree of organizational complexity than the random graphs.

\begin{figure*}
\centering
\includegraphics[scale = 0.37,angle=0.00, clip=true]{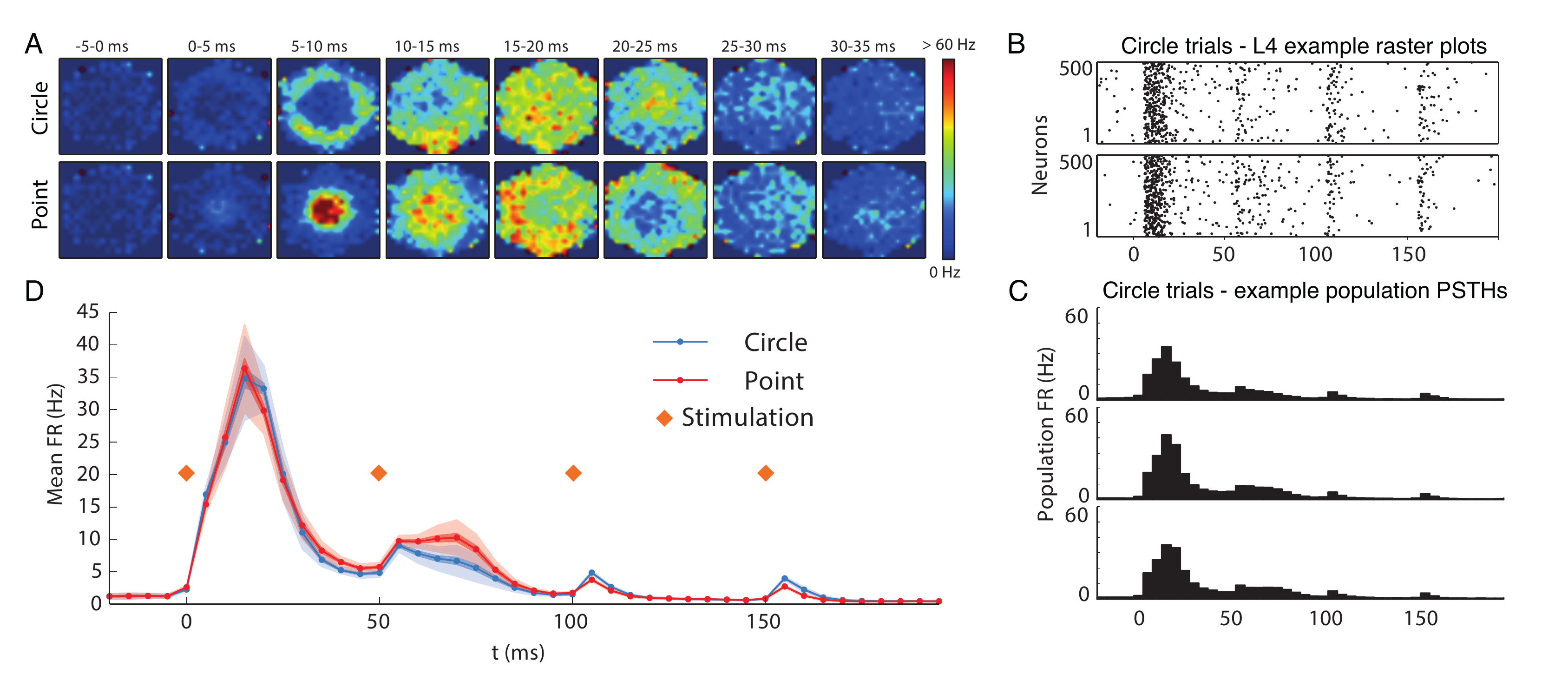}
\caption{
\emph{(A)} Average firing rate (top-down projection) in the stimulated microcircuit, plotted during the first 35 ms after the first stimulation at t=0 ms in the {Point vs.~Circle} experiment. \emph{(B)} Raster plots of the same 500 neurons randomly picked from layer 4, for two trials of the circle stimulus. \emph{(C)} Population PSTH of all neurons in the microcircuit for three trials of the Circle stimulus. \emph{(D)} Mean firing rate of the Circle and Point stimuli, between $t$ and $t+5$ ms, where light shading indicates the standard deviation and dark shading the error of the mean.
}
\label{fig:figure3}
\end{figure*}

Topological methods also enabled us to distinguish functional responses to different input patterns fed into the microcircuit through thalamo-cortical connections.  We ran simulations of neural activity in one of the reconstructed microcircuits during one second, over the course of which a given stimulus was applied every 50 ms (Figure \ref{fig:figure3}). We then binned the output of the simulations by 5 ms timesteps and associated to each timestep a \emph{transmission-response graph}, the vertices of which are all of the neurons in the microcircuit and the edges of which encode connections in the microcircuit whose activity in that time step leads to firing of the postsynaptic neuron (Figure \ref{fig:figure4}). The size of the time bins and the precise rule for formation of the transmission-response graph for each time bin are biologically motivated, as explained in more detail in the Supplementary Methods section (SI, Supplementary Methods, SM1).  

From the time series of transmission-response graphs for each of 20 trials of two different stimuli (called Circle and Point for geometric reasons (Figure \ref{fig:figure4}A), we derived time series of two non-topological metrics (mean firing rate and number of edges in the transmission-response graph) and five topological metrics (the number of $3$-cliques, EC, $\beta_{0}$, $\beta_{1}$, and $\beta_{2}$) and applied a Gaussian  Bayes classifier (SI, Supplementary Methods, SM2) to determine how successfully each of the metrics classified the 40 trials in the time bins corresponding to the first two stimulations and in the time bins immediately following those stimulations (Figure \ref{fig:figure5}). In each of those crucial time bins, the metrics that were most successful at classification the number of $3$-cliques (denoted 2D in the figure), $\beta_{2}$, and, in one case, the Euler characteristic (Figure \ref{fig:figure2}A).

We expect the methods applied here will prove useful for studying networks in general.

\section{Structural topology}
 We computed the binary adjacency matrices of all 42 digitally reconstructed microcircuits and then generated the associated \hadgesh{directed flag complexes} (SI, Supplementary Text, ST1.2), which are oriented simplicial complexes  encoding the connectivity of all orders of the underlying directed graph:  to each directed $n$-clique (SI, Supplementary Text, ST1.2) in the underlying graph corresponds to an oriented $(n-1)$-simplex in the flag complex, and the faces of a simplex correspond to the directed subcliques of its associated directed clique (Figure \ref{fig:figure1} B and C). For each neuron in the microcircuit, there is a vertex in the underlying directed graph that is labelled with the unique \emph{global identification number (GID)} of the neuron.  The $(j,k)$-coefficient of the structural adjacency matrix is 1 if and only if there is a directed connection in the microcircuit from the neuron with GID $j$ to the neuron with GID $k$.   We refer to this adjacency matrix as the \hadgesh{structural matrix} of the microcircuit and to its associated directed flag complex as a \hadgesh{neocortical microcircuit complex} or \hadgesh{N-complex}.

Having computed each of the 42 N-complexes, we counted the simplices in each dimension. For comparison with non-biological matrices, we generated five Erd\H os-R\'enyi random graphs \cite{ER} of a comparable size (31,000 vertices) and connection probability $0.8\%$, the same as the average arising from the structural matrices of the microcircuits  (SI, Supplementary Methods SM3.1). We refer to the associated directed flag complexes as  \hadgesh{ER-complexes}.  

To have a more biological control, we also generated 20 adjacency matrices, given by partly randomizing the structural matrix of one of the average microcircuits, taking into account its biologically meaningful division into six layers in 10 cases and into 55 morphological neuron types (m-types) \cite{BBP-ref} in 10 cases. The randomization was carried out so that the distance-dependent connection probability for all pairs of layers (respectively, pairs of m-types) was identical to that of the original matrix, i.e., for each pair of layers (respectively, m-types) the number of connections between them was the same as that of  the original and for each $25\, \mu\text{m}$ distance bin the number of connections was identical. The matrices are completely random otherwise (SI, Supplementary Methods SM3.2, SM3.3).  We call the associated directed flag complexes \emph{L-complexes} (respectively, \emph{M-complexes}).  Note that since each m-type is restricted to a fixed layer, the M-complex should retain more of the structure of the original N-complex than the L-complex.   Our final and most biological control consisted in the generation of 10 connectivity matrices for 31,000 neurons according to Peters' Rule \cite{Peters1}, \cite {Peters2} for which the associated directed flag complexes are called \emph{P-complexes} (SI, Supplementary Methods SM3.4). Having carried out the computations for 10 control matrices out of each randomized set of 20, the very small variance in the results convinced us  that no further computations should be needed.

The resulting distribution of simplices displayed highly consistent behavior among the N-complexes, all of which we computed, with a small variation among the samples arising from different rats. Note that Figure \ref{fig:figure1} represents the analysis only of the seven N-complexes arising from the average reconstruction because the randomizations are based on those microcircuits. The ER-complexes showed almost identical behavior among the different instances, as did the L-complexes, M-complexes, and P-complexes. On the other hand, the  N-complexes exhibited remarkably different distributions from the various random complexes (Figure \ref{fig:figure1} D), with much greater numbers of simplices and simplices of significantly higher dimension.  We computed the Euler characteristic  of all N-complexes, as well as that of the various random complexes, obtaining large positive values in all cases, due to the predominance of even-dimensional (particularly 2-dimensional) simplices.

The Betti numbers (SI, Supplementary Text, ST1.2) of a  simplicial complex provide a much finer and more sophisticated measure of its organizational complexity than the dimension-wise simplex count or the Euler characteristic.  The $n$-th Betti number, $\beta_{n}$, counts the number of chains of simplices intersecting along faces to create an ``$n$-dimensional hole'' in the complex, which requires a certain degree of organization among the simplices.  On the other hand, computation of the Betti numbers is much more expensive than  that of  the directed flag complex of a directed graph or its Euler characteristic. In fact, the sheer size of the complexes we considered here made it practically impossible to do so on a computer with 256 GB of RAM. We succeeded in computing the highest nonzero Betti numbers of the N-complexes, however, by restricting our attention to the 5-th and 6-th coskeleta (SI, Supplementary Text, ST1.2). The top Betti number in all N-complexes appeared in dimension 5, with $\beta_{5}$ varying between 1 and 80 (Figure \ref{fig:figure2}D). By contrast, $\beta_{n}=0$ for all $n>3$ for all ER-complexes and P-complexes considered, while $\beta_{n}=0$ for all $n>4$ for all L-complexes and M-complexes.  Moreover $\beta_{4}$ varies between 0 and 6 for all L-complexes and M-complexes, so that these Betti numbers are almost negligible.

\section{Functional topology}
We  tested our methods on active microcircuits as well. In an experiment that we call the  \hadgesh{Point vs.~Circle test}, we activated in a simulation the incoming  thalamo-cortical fibers of one of the average   that the stimulated fibers formed first a point shape, then a circle shape \cite{BBP-ref}. 
The size of the point shape was chosen such that the average firing rate of the neurons was essentially the same as for the circle shape, and in both cases the fibers were activated regularly and synchronously with a frequency of 20 Hz for one second, similar to the whisker deflection approximation in  \cite[Figure 17A]{BBP-ref}.  We performed 20 trials of each stimulus (Figure \ref{fig:figure3}).  The trials of each stimulus  exhibit biological trial-to-trial variability in the neural response, due to the stochasticity of the synapse models and of some of the ion channel models. The aim of this experiment was to determine whether our topological methods were able to classify the two different stimuli, the point and the circle better than the firing rate, which is largely overlapping for the first two stimulations (see Figure \ref{fig:figure3}D).

After a systematic analysis to determine the appropriate time bin size and conditions for probable spike transmission from one neuron to another (SI, Supplementary Methods, SM1.4), we divided the activity of the microcircuit into 5 ms time bins for 1 second after the initial stimulation and recorded for each $0\leq n<200$ a functional connectivity matrix $A(n)$ for the times between $5n \text{ ms}$ and $5(n+1) \text{ ms}$.   The $(j,k)$-coefficient of the binary matrix $A(n)$ is $1$  if and only if the following three conditions are satisfied, where $s_i^j$ denotes the time of the $i$-th spike of neuron $j$.
\begin{enumerate}
\item[(1)] The $(j,k)$-coefficient of  the structural matrix is 1, i.e., there is a structural connection from the neuron with GID $j$ to the neuron with GID $k$.
\item[(2)] There is some $i$ such that  $5n \text{ ms}\leq s_{i}^{j}<5(n+1) \text{ ms}$, i.e., the neuron with GID $j$ spikes in the $n$-th time bin.
\item[(3)] There is some $l$ such that $0 \text{ ms}<s_{l}^{k}-s_{i}^{j}< 7.5 \text{ ms}$, i.e., the neuron with GID $k$ spikes after the neuron with GID $j$, within a 7.5 ms interval.
\end{enumerate}
We call the matrices $A(n)$ \hadgesh{transmission-response matrices}, as  it is reasonable to assume that the spiking of neuron $k$ is influenced by the spiking of neuron $j$ under conditions (1)--(3) above.   

\begin{figure*}
\centering
\includegraphics[scale = 0.4,angle=0.00, clip=true]{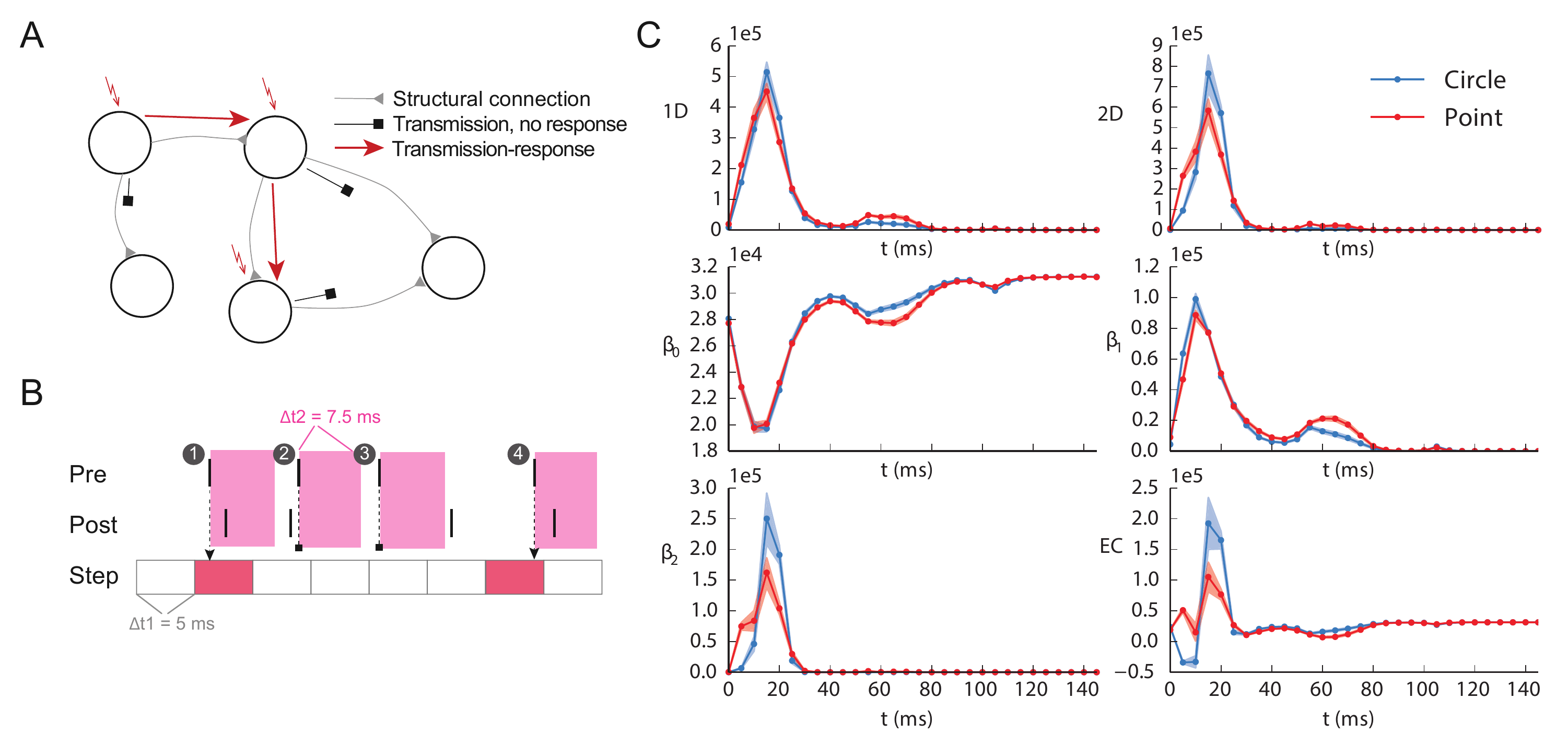}
\caption{
\emph{(A)} Schematic representation of the transmission-response paradigm:  there will be an edge from $j$ to $k$ in the graph associated to particular time bin if and only if there is a physical connection from neuron $j$ to neuron $k$, neuron $j$ fires in the time bin, and neuron $k$ fires at most 7.5 ms after the firing of neuron $j$. Here, shading indicates the error of the mean.\emph{(B)} Schematic representation of those firing patterns involving a presynaptic and a postsynaptic neuron that lead to an edge in the transmission-response graph, with a red block indicating successful transmission and a white block indicating lack of transmission. \emph{(C)}  Time series plots of the average value of the metrics 1D (number of 1-simplices), 2D (number of 2-simplices), $\beta_{0}$ (the zeroth Betti number, i.e., the number of connected components), $\beta_{1}$ (the first Betti number), $\beta_{2}$ (the second Betti number), and EC (the Euler characteristic) for the Circle and Point stimuli. Here, shading indicates the error of the mean.
}
\label{fig:figure4}
\end{figure*}

The goal of  the \hadgesh{Point vs.~Circle test} was to determine whether topological metrics, such as simplex counts, Betti numbers and Euler characteristic, could classify correctly two groups of stimuli of a similar nature and whether these metrics contain more information than the mean firing rate.   In Figure \ref{fig:figure4}C we provide plots of the time series of the average zeroth, first, and second Betti numbers, of the average numbers of 1- and 2-simplices, and of the average Euler characteristic for 20 trials of each stimulus.  We applied a Gaussian  Bayes classifier (SI, Supplementary Methods, SM2) to each metric in each time bin, to determine their success rate at classifying the various trials of the stimuli.  To compare, we also classified the stimuli according to the mean firing rates. To allow for a fair comparison, we used three mean firing rates (between $t$ to $t+5$, $t+5$ to $t+10$, and $t+10$ to $t+15$ ms) for the classification at each time step $t$, since the transmission-response edges for time step $t$ are based on information from up to $t+12.5$ ms.

\begin{figure}[h!]
\centering
\includegraphics[scale = 0.45,angle=0.00, clip=true]{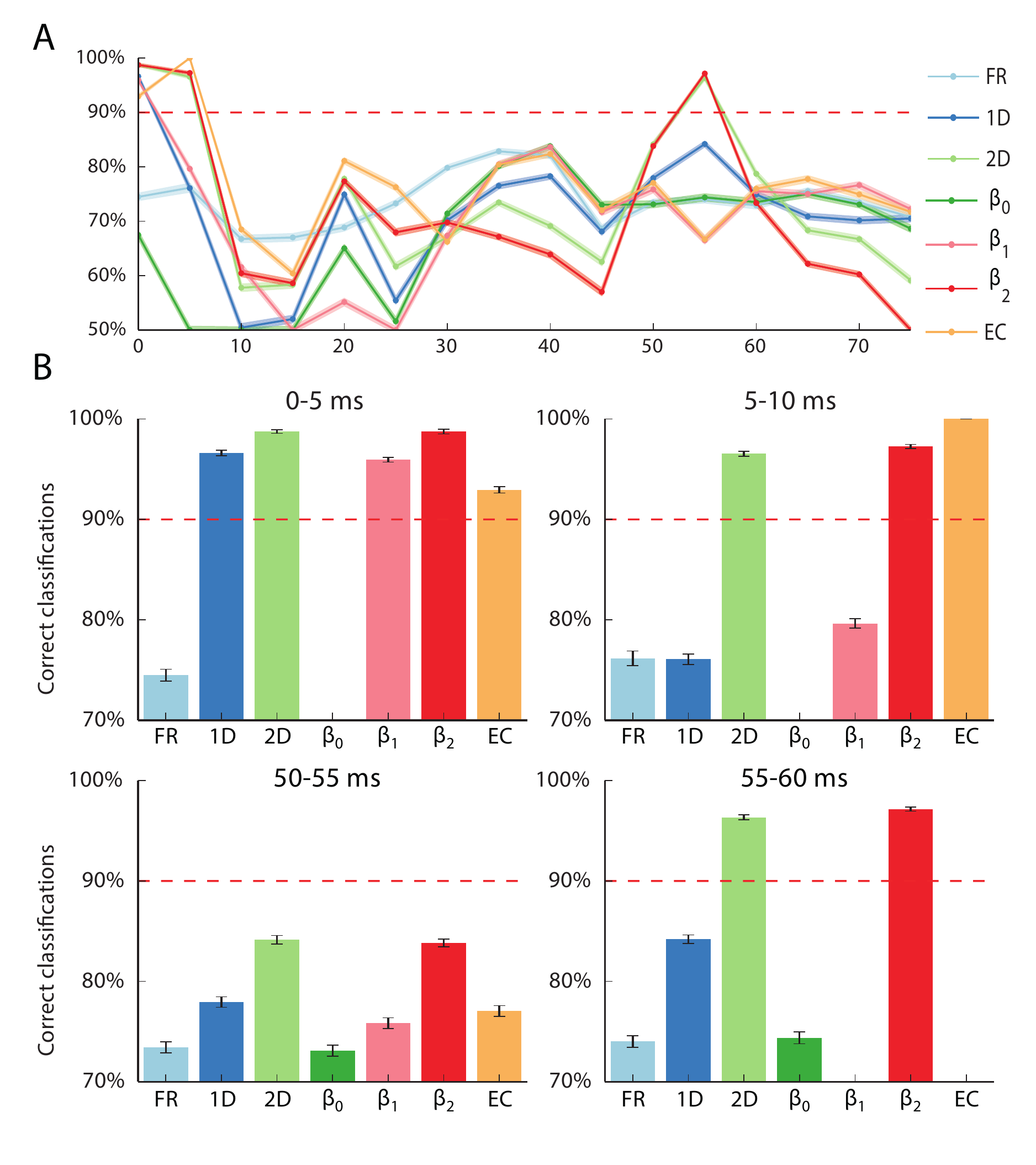}
\caption{\emph{(A)} Times series plot for the first 80 ms of the 40 trials of the percentage of correct classifications performed by a Gaussian  Bayes classifier based on each of the metrics FR (sequences of mean firing rates over three consecutive time bins), 1D (number of 1-simplices), 2D (number of 2-simplices), $\beta_{0}$ (the zeroth Betti number, i.e., the number of connected components), $\beta_{1}$ (the first Betti number), $\beta_{2}$ (the second Betti number), and EC (the Euler characteristic). \emph{(B)} Graphs depicting the percentage of correct classifications performed by a Gaussian  Bayes classifier based on each of the metrics in four particularly important time bins:  from 0 to 5 ms (immediately after the initial stimulation), from 5 to 10 ms, from 50 to 55 ms (the time bin immediately after the second stimulation), and from 55 to 60 ms. 
}\label{fig:figure5}
\end{figure}

As illustrated by Figure \ref{fig:figure5} A, none of the metrics considered, topological or otherwise, succeeded very well at classifying the stimuli for times between 10 ms and 50 ms after the initial stimulation, which is not surprising given the strong similarity between the spatial propagation of activity of the two stimuli during this period (Figure \ref{fig:figure3}).  On the other hand, in the very first time bin, immediately after the initial stimulation, the 1- and 2-dimensional simplex counts and $\beta_{1}$ and $\beta_{2}$ all classify very well.  In the second time bin the 2-dimensional simplex count and $\beta_{2}$ continue to classify very well, and the Euler characteristic classifies even better. Immediately after the second stimulation, from 50 ms to 55 ms after the initial stimulation, none of the metrics performs very well, but the  2-dimensional simplex count and $\beta_{2}$ still have the highest success rate.   In the next time bin, from 55 ms to 60 ms after the initial stimulation,  the  2-dimensional simplex count and $\beta_{2}$ again classify very well and are the only metrics to do so.
In all of these cases, the topological metrics far outperform the metric based on firing rate.

\section{Discussion}
We have introduced topological analysis of directed graphs encoding structural or functional connectivity of digital reconstructions of neural microcircuits.  We showed in particular that these directed graphs differed significantly from random graphs of both Erd\H os-R\'enyi-type and types taking into account biologically constrained, distance-dependent connection probabilities.  The topological analysis revealed not only the existence of high-dimensional simplices representing the most extreme form of circuit ``motifs'' - all-to-all connectivity within a set of neurons - that have so far been been reported for brain tissue, but also that there are a surprisingly huge number of these structures. We established moreover that topological methods effectively distinguish functional responses to distinct thalamic stimuli, introducing a new measure of the spatio-temporal activity responses generated by neural tissue.  The results of our topological analysis of biologically realistic digital reconstructions provide a convincing argument for considering topology as a useful mathematical tool for analyzing the structural and functional connectome of neural circuits.

Our results lead naturally to many new questions, most notably concerning the biological significance of the high-dimensional simplices and homology classes we have discovered in the digitally reconstructed neocortical microcircuits. We intend to explore these questions in future studies. In particular we hypothesize that the time series of different topological metrics could reveal an evolving spatio-temporal code that goes beyond either rate or timing information to one that incorporates the structural organization. Such metrics could yield a deeper understanding of how the structural organization constrains emergent functional states. Age-dependent changes in such digital reconstructions may help reveal even more complex topological structures with development, and changes introduced by synaptic plasticity may reveal structures associated with learning and memory.

We expect the topological approach to studying directed graphs that we implement here will also prove useful in applications of network science outside of neuroscience, in the study of networks exhibiting intricate directed connectivity patterns, such as gene and protein networks, VLSI circuits, and electrical grids. The obvious utility of the directed flag complex in these applications may also encourage theorists to establish results analogous to those established by Kahle concerning Betti numbers of undirected flag complexes of random graphs \cite{kahle}.

\section{Materials and methods}
\subsection{Computation of flag complexes and their Betti numbers} We represent the directed flag complex of a directed graph by a reference-based data structure, using vectors to store the references to the simplices in the simplicial complex. The required storage space grows linearly with the number of vertices and with the number of edges. A publicly available C++ implementation of the code will be available on http://neurotop.gforge.inria.fr/. All homology computations carried out for this paper were made with $\mathbb{F}_2$ coefficients, using the boundary matrix reduced by an algorithm from the PHAT~\cite{phatURL} library.   For further details, please see (SI, Supplementary Text, ST2).

\subsection{The Point vs.~Circle experiment}
The stimulated reconstructed microcircuit is innervated by 310 VPM fibers, whose horizontal centers of innervation are evenly distributed over the microcircuit (one fiber per mini-column). It is therefore possible to activate the microcircuit with topographically different stimuli by selecting only a subset of these 310 fibers. Here we used two different stimuli, a point and a circle, which were calibrated by adjusting the respective number of fibers to evoke an overall similar mean firing rate (i.e., close enough to prevent clearly distinguishing between the two stimuli simply by the mean population firing rate). The microcircuit was stimulated by synchronous spikes, similar to the whisker deflection experiment described by Markram et al. (2015). The point stimulus consisted of synchronous spikes in the 46 neighboring fibers of  the center of the microcircuit, whereas the circle stimulus involved 56 fibers near the periphery of the microcircuit. The stimulation was repeated every 50 ms, but only the firing rates after the first two stimulations (at 0 and 50 ms) are overlapping.

We used a Gaussian na\"ive Bayes classifier \cite{scikit-learn}, where we performed 500 classification trials, randomly choosing 15 trials of each stimulus to be part of the training data, and five trials of each stimulus to be part of the test data. We then obtained the mean ratio of successfully classified test data points using 500 different training and test sets. The classification of the firing rate used the firing rates of three consecutive time bins, to make it a fairer comparison, since the edges may contain firing rate information of more than two time bins, over a range of 12.5 ms.

\subsection{Computation of transmission-response matrices}
Transmission-response matrices were calculated according to the specifications mentioned above, using a custom-written program in the Python programming language. It combined the matrix of synaptic connections (structural matrix), constructed as part of the standard reconstruction process of the BBP, with the spiking output of a simulation run and user-defined values for time steps $\Delta t_1$ and $\Delta t_2$ (5 and 7.5 ms in our analyses). For further details, please see (SI, Supplementary Methods ,SM1).

\subsection{Gaussian Bayes classifiers}
The Gaussian Bayes classifier minimises the probability of misclassification under the assumption that the distributions are Gaussian. We randomly split the data  into training and testing sets. Using the training set we model the distributions of the dot and circle classes by Gaussians 
$\mathcal{N}(\hat{\mu}_{\text{dot}}, \hat{\sigma}_{\text{dot}}^2)$ and 
$\mathcal{N}(\hat{\mu}_{\text{circle}}, \hat{\sigma}_{\text{circle}}^2)$ respectively. 
Assuming a uniform prior and Gaussian distributions, Bayes' theorem provides a classifier
$$\text{Class}(x)=
\argmax_{c\in \{\text{dot}, \text{circle}\}}􏰄
\frac{1}{\sqrt{2\pi\hat{\sigma}_c^2}}
\exp􏰂\left(\frac{-(x-\hat{\mu}_c)^2}{2\hat{\sigma}_c^2} \right) .$$

\section{Acknowledgments}
This work was supported by funding from the ETH Domain for the Blue Brain
Project (BBP). The BlueBrain IV IBM BlueGene/Q system is financed by ETH
Board Funding to the Blue Brain Project and hosted at the Swiss National
Supercomputing Center (CSCS).
We thank Ahmet Bilgili for providing the visualization of the microcircuit in Figure \ref{fig:figure1}. Partial support for P.D. was provided by the GUDHI project, supported by an Advanced Investigator Grant of the European Research Council and hosted by INRIA.  M.S. was supported by the NCCR Synapsy of the Swiss National Science Foundation.

\vfill\eject

\newcommand{\beginsupplement}{%
        \setcounter{table}{0}
        \renewcommand{\thetable}{S\arabic{table}}%
        \setcounter{figure}{0}
        \renewcommand{\thefigure}{S\arabic{figure}}%
     }

\beginsupplement

\begin{center}
{\Large SI Appendix}\\
To accompany ``Topological analysis of the connectome of digital reconstructions of neural microcircuits."\\
Pawe\l\ D\l otko, Kathryn Hess, Ran Levi, Max Nolte, Michael Reimann, Martina Scolamiero, Katharine Turner, Eilif Muller, Henry Markram
\end{center}
\bigskip

\section*{Contents}
\begin{enumerate}
\item Supplementary Methods (3 pages) 
\item Supplementary Text (4 pages)
\item Supplementary Figures (5 pages)
\end{enumerate}

\section*{Supplementary Methods}

\section*{SM1. Optimization of the parameters for the transmission-response matrices}

The transmission-response matrices that allow us to analyze activity in an experiment  (cf.~the section on Functional Topology in the main body of the article)  form a sequence  depending on two parameters, $\Delta t_{1}$ and $\Delta t_{2}$. The number of matrices in the sequence is the duration of the experiment divided by $\Delta t_1$. In other words for a given experiment of duration $T$ and fixed $\Delta t_i$, we obtain a sequence of matrices $S(\Delta t_1, \Delta t_2) = \{A(n)=A(n,\Delta t_1, \Delta t_2)\}_{n=1}^{N}$, where $N$ is the integer value of $T/\Delta t_1$.

For fixed values of $\Delta t_{1}$ and $\Delta t_{2}$, the corresponding sequence $\{A(n)\}_{n=1}^{N}$  is obtained as follows. The spiking output of the simulation is first converted into lists of spike times, one for each neuron. Standard histogram methods, binning by $\Delta t_1$, are applied to each list to determine in which time steps a presynaptic neuron fired. For each time bin in which a particular neuron fired, the exact timing of its first spike in that bin is then compared to the full list of spike times of each neuron it innervates, to ascertain which of them had spiked at most $\Delta t_2$ ms after the presynaptic neuron. (Spiking of a pair of neurons within $\Delta t_2$ ms is ignored if they are not structurally connected.) For all pre-postsynaptic pairs satisfying this constraint on spike timing, the corresponding entry in the transmission-response matrix for that time step is set to 1 and all others to 0. More precisely, the $(j,k)$-coefficient of the binary transmission-response matrix $A(n)$ corresponding to the $n$-th time bin is $1$  if and only if the following three conditions are satisfied, where $s_i^j$ denotes the time of the $i$-th spike of neuron $j$.
\begin{enumerate}
\item[(1)] The $(j,k)$-coefficient of  the structural matrix is 1, i.e., there is a structural connection from the neuron with GID $j$ to the neuron with GID $k$, so that they form a pre-post synaptic pair.
\item[(2)] There is some $i$ such that  $n\Delta t_1 \text{ ms}\leq s_{i}^{j}<(n+1)\Delta t_1 \text{ ms}$, i.e., the neuron with GID $j$ spikes in the $n$-th time bin.
\item[(3)] There is some $l$ such that $0 \text{ ms}<s_{l}^{k}-s_{i}^{j}< \Delta t_2 \text{ ms}$, i.e., the neuron with GID $k$ spikes after the neuron with GID $j$, within a $\Delta t_2$ ms interval.
\end{enumerate}

Starting with firing data from spontaneous activity in the reconstructed microcircuit, we generated sequences of 20 transmission-response matrices for $\Delta t_i\in \{1 ,2 ,5 ,10, 20, 50, 100\}$ ms, thus creating 49 such sequences corresponding to every possible choice of the pair $(\Delta t_1, \Delta t_2)$.  We refer to each of these sequences as the \emph{true transmission-response matrices} corresponding to the pair $(\Delta t_1, \Delta t_2)$. 

In the rest of this section, we describe the procedure for optimizing the choice of the time intervals $\Delta t_{1}$ and $\Delta t_{2}$ so that the associated true transmission-response matrices best reflect the actual successful transmission of signals between neurons in the microcircuit.

\subsection*{SM1.1. Properties of the transmission-response matrix}
The nonzero coefficients in a transmission-response matrix are a subset of those in the structural matrix. Due to the partly stochastic behavior of the \emph{in silico} microcircuit, the subset will vary even for subsequent applications of the same stimulus. In fact, even an exact repetition of the same conditions will lead to different transmission-response matrices, if the random number generator is seeded differently. It follows that the generation of the transmission-response matrices for a given stimulus should be considered as a stochastic process. With the correct choice of the parameters $\Delta t_i$, the matrices should reflect how the microcircuit  processes a stimulus and thus take into account parameters of neural processing, such as pre-post synaptic interaction. 

To find parameters $\Delta t_{1}$ and $\Delta t_{2}$ that maximize the degree to which neural processing is captured by the transmission-response matrices, we first develop a stochastic model for synaptic firing that takes into account neural processing and that depends on $\Delta t_{1}$ and $\Delta t_{2}$. For the purpose of this analysis, we assume that the true transmission-response matrices are compatible with this model. 

Based upon our model for synaptic firing, we formulate a simplified model that ignores neural processing.  For this simplified model and for any choice of parameters $\Delta t_{1}$ and $\Delta t_{2}$, we explain how to obtain transmission-response matrices from actual firing data, by shuffling the firing data appropriately, then applying the algorithm for generating a transmission-response matrix of the previous section.  Finally, for each choice of the parameters $\Delta t_1$ and $\Delta t_2$, we compare the true transmission-response matrices for spontaneous activity in the reconstructed microcircuit to those obtained by the simplified generation process.  The parameters that we work with in the main body of the paper are the $\Delta t_1$ and $\Delta t_2$ that maximize the difference (measured by the ratio of the numbers of ones in the matrices) between the actual transmission-response matrices and those resulting from the simplified model. 

\subsection*{SM1.2. Stochastic model with neural processing}
Fix time intervals $\Delta t_{1}$ and $\Delta t_{2}$. Let $A = (a_{ij})$ denote the structural  matrix of a reconstructed microcircuit, and let $A(n) = (a^n_{ij})$ denote the transmission-response matrix of the $n$-th time bin, based on firing data from a trial of simulated activity in the microcircuit, for the given intervals $\Delta t_{1}$ and $\Delta t_{2}$. By Condition (1) above, if $a^n_{ij}=1$ for any $n$, then $a_{ij} = 1$.  It is reasonable to consider $A$ to be static, at least over the time periods considered here.   

We want to compute the probability that $a^n_{ij} = 1$, given that $a_{ij} = 1$, so we need to determine on  which parameters and properties this probability depends.
According to the definition of transmission-response matrices, a presynaptic and a postsynaptic spike are required for $a^n_{ij}$ to be 1. To simplify the analysis somewhat, we assume  that each neuron $n_i$ has a \emph{time-dependent, instantaneous firing rate} $F^i(t)$ that determines spiking probability at time $t$, i.e., spiking can be described as an inhomogeneous Poisson process. Under this assumption,   the expected number $m_{\Delta t_{1}}^i(t_{0})$ of spikes of neuron $n_i$ in the  interval $[t_{0}, t_{0}+\Delta t_{1}]$ can be computed as 
\[m_{\Delta t_{1}}^i(t_{0}) = \int_{t_{0}}^{t_{0} + \Delta t_{1}}F^i(u)du.\]
If $K_{\Delta t_1}^i(t_0)$ denotes the probability that neuron $n_i$ spikes at least once in the interval $[t_{0}, t_{0}+\Delta t_{1}]$, then 
$$
K_{\Delta t_1}^i(t_{0}) = 1 - \mathcal{P}\big(m_{\Delta t_1}^i(t_{0})\big) = 1 - e^{-m_{\Delta t_1}^i(t_{0})},
$$
where $\mathcal{P}(\lambda)$ is the Poisson probability mass function with parameter $\lambda$ at 0. (Recall that if $X$ is a random variable that counts the number of spikes of neuron $n_i$ in the interval $[t_0, t_0+\Delta t_1]$, then $\mathcal{P}\big(m_{\Delta t_1}^i(t_{0})\big)$ is the probability  that $X=0$.) If the change in $F^i(t)$ is slow compared to $\Delta t_1$, then $m_{\Delta t_1}^i(t) \approx F^i(t) \cdot \Delta t_1$. Moreover,   $1 - \mathcal{P}(\lambda) \approx \lambda$ for small values of $\lambda$. For small enough  $\Delta {t_{1}}$,  the expected number $m_{\Delta t_{1}}^i(t_{0})$ of spikes of neuron $n_{i}$ will certainly be small, and change in $F^i(t)$ will be slow in compared to $\Delta t_1$, so that we may assume that
$$K_{\Delta t_1}^i(t_{0}) \approx F^i(t_{0}) \cdot \Delta t_1.$$

For the postsynaptic spike the situation is  more complicated. As there is a causal relation between presynaptic and postsynaptic firing, mediated by synaptic transmission, we need to estimate the  conditional probability of at least one postsynaptic spike, given that at least one presynaptic spike occured. Let $n_{i}$ and $n_{j}$ denote neurons such that $a_{ij}=1$. Let $s_0\in [t_0, t_0+\Delta t_1]$ denote  the time of the first presynaptic spike in this interval. 
Let $X_{\Delta t_2}^j(s_0)$ denote the random variable whose value is the number of times  neuron $n_j$ spiked  in the time window $[s_0, s_0+\Delta t_2]$. 
Let $Y_{\Delta t_1}^{i}(t_0)$ denote the random variable whose value is the number of times neuron $n_i$ spiked in the time interval $[t_0, t_0 + \Delta t_1]$.  We need to estimate the conditional probability 
\[P\big(X_{\Delta t_2}^j(s_0)>0\,|\,Y_{\Delta t_{1}}^{i}(t_0)>0\big).\]

The nature of this interaction is very intricate and depends on the identities of the presynaptic and postsynaptic neurons, the spiking history of the presynaptic neuron before $s_0$, and all other synaptic input the postsynaptic neuron received. It can be described as governed by some  function $G^{ij}$ modulating the spiking probability of the postsynaptic neuron $n_j$. This function takes as parameters the expected number of spikes of neuron $n_j$ in the interval $[s_0, s_0+\Delta t_2]$, the time $t_0$, and the ``spiking history" of the presynaptic neuron $n_i$ until $s_0$, which we write as a function $s_*^i(t)$ evaluated at $s_0$, giving rise to the expression
$$
P\big(X_{\Delta t_2}^j(s_0)>0\,|\,Y_{\Delta t_{1}}^{i}(t_0)>0\big) = 1 - e^{-G^{ij}(m_{\Delta t_2}^j(s_{0}), t_0, s_*^i(s_0))}.
$$

Summarizing the analysis above, the following formula provides a good estimate of the probability that $a^n_{ij} = 1$ if  $a_{ij} = 1$, for small enough $\Delta t_{1}$ and $\Delta t_{2}$, where $s_{0}$ denotes the time of the first presynaptic spike in the interval $\left[n \Delta t_{1}, (n+1) \Delta t_{1}\right]$ and $t_{0}=n \Delta t_{1}$.
\begin{equation}\label{eqn:model}
\begin{aligned}
P\left(a^n_{ij} = 1|a_{ij} = 1\right) &= \left(1 - e^{-m_{\Delta t_1}^i(t_{0})}\right) \cdot \left(1 - e^{-G^{ij}(m_{\Delta t_2}^j(s_{0}), t_0, s_*^i(s_0))}\right)\\
&\approx F^i(t_{0}) \cdot \Delta t_1\cdot G^{ij}\big(F^j(s_{0}) \cdot \Delta t_2, t_{0}, s_*^i(s_0)\big).
\end{aligned}
\end{equation}
This conditional probability encodes not only the distinctive features of the structural connectivity (via $a_{ij}$) but also the potentially stimulus-dependant neuron-specific firing rates (via $F^i$ and $F^j$) and their co-variation. Most crucially, it captures  the stimulus-dependent functional modulation of postsynaptic firing by a presynaptic spike as well.   We assume that the true transmission-reponse matrices  capture the actual transmission of spikes according to the model of synaptic firing described by this formula.

\subsection*{SM.1.3. Null hypotheses: no neural processing}
We introduce here a simplified model of synaptic spiking that  is based upon formula [\ref{eqn:model}] but that ignores pre-post synaptic interaction. We then explain how to obtain transmission-response matrices that correspond to this simplified model from firing data arising from simulated activity. 

We begin by setting each  $G^{ij}$ to be the projection onto the first component, ignoring the pre-post synaptic interaction. After this simplification, the approximation obtained in the previous section now reads
\[P(a^n_{ij} = 1|a_{ij}=1)\approx F^i(t_{0}) \cdot F^j(s_{0}) \cdot \Delta t_1\cdot \Delta t_2,\]
where $s_{0}$ denotes the time of the first presynaptic spike in the interval $\left[n \Delta t_{1}, (n+1) \Delta t_{1}\right]$ and $t_{0}=n \Delta t_{1}$, as before.  Since this drastic simplification neglects the central aspect of neural computation - pre-post synaptic interaction - it gives rise to control cases for each pair of parameters $(\Delta t_1, \Delta t_2)$ and each choice of firing rate functions $F^i(t)$. Comparison of the true transmission-response matrices for each pair of parameters  to the corresponding control matrices for the same pair and a specific choice of the functions $F^i(t)$ will allow us  to determine values for $\Delta t_1$ and $\Delta t_2$ for which the true transmission-response matrix optimally reflects neural processing.

We assume moreover that  the individual firing rates consist of a neuron-dependent frequency that is up- or down-regulated by a global time series, i.e., that  $F^i(t) = f(i) \cdot F(t)$, for some function $F(t)$ and some constant $f(i)$ for each neuron $n_{i}$. Transmission-response matrices corresponding to this simplified model for fixed $\Delta t_{1}$ and $\Delta t_{2}$, which we call \emph{simplified transmission-response matrices},  can be generated  by first shuffling all recorded spikes from simulated activity in the reconstructed microcircuit, while preserving both the number of spikes per neuron and per time bin, then applying the usual transmission-response matrix generation method.

\subsection*{SM.1.4. Optimization of parameters}
The difference between the true transmission-response matrices and  the control case described above is a consequence of the pre-post synaptic interaction. Comparison with the control case enables us therefore to measure how well that interaction is captured in the true transmission-response matrices. In particular, it is reasonable to optimize the parameters $\Delta t_1$ and $\Delta t_2$ so that the difference between the true transmission response matrices arising from actual simulation data and those arising in the control cases is maximized, as a maximal difference indicates that the effect of the pre-post synaptic interaction is captured optimally by the true transmission-response matrices. 

The comparison between the true transmission-response matrices and the control cases was carried out by first producing 20 true transmission-response matrices and 20 simplified transmission-response matrices based on firing data obtained from spontaneous activity in the reconstructed microcircuit for every pair $(\Delta t_{1}, \Delta t_{2})$, where $\Delta t_i\in \{1 ,2 ,5 ,10, 20, 50, 100\}$ ms for $i=1,2$.  The number of ones in each matrix was then computed and the average taken over each set of 20 matrices.    Since no stimulus was applied to the microcircuit, the averages computed are meaningful, since the firing data should be fairly homogeneous across the time bins.

\begin{figure*}[!htp]
\centering
\includegraphics[scale=0.45,angle=0.00, clip=true]{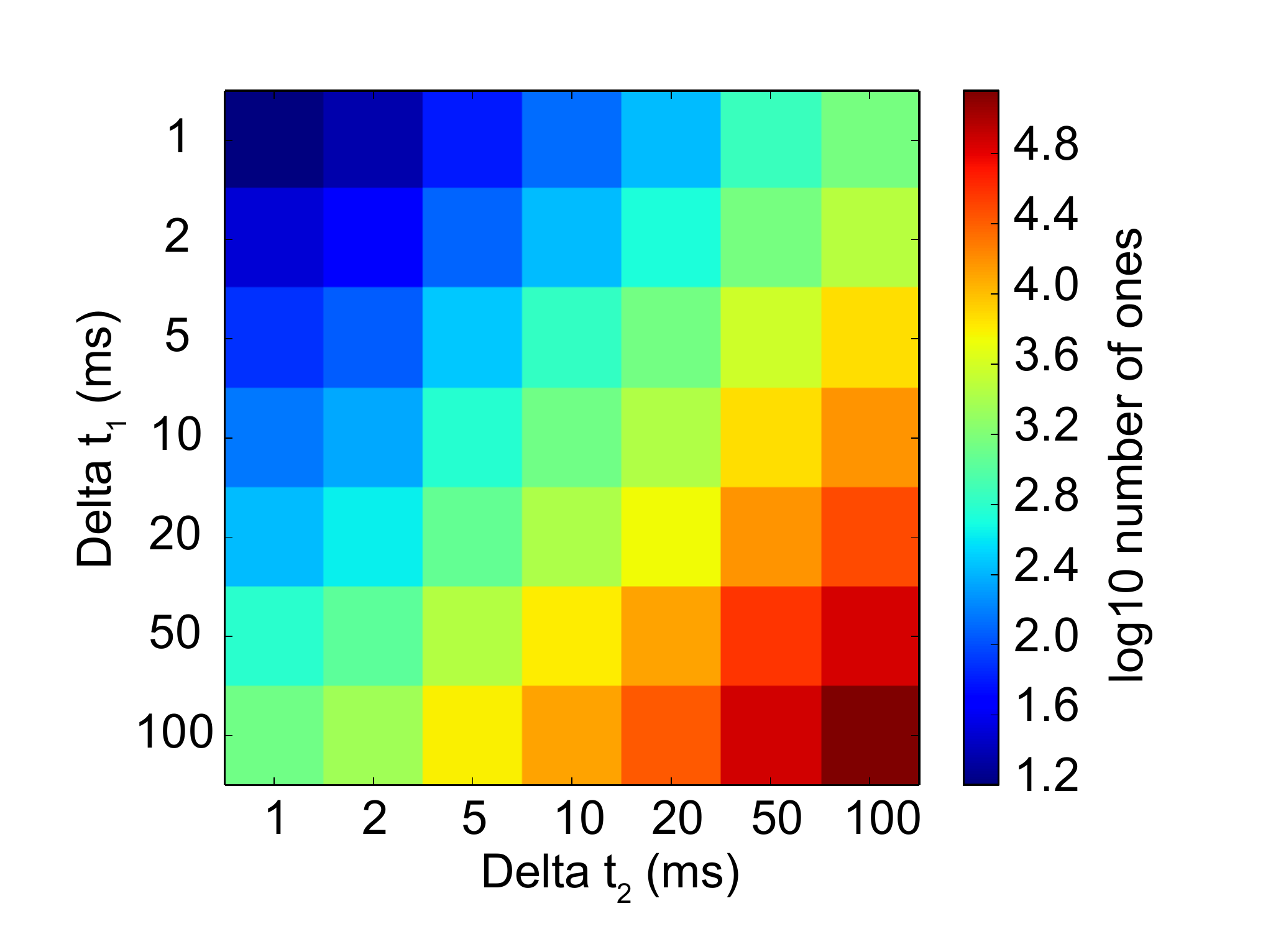}
\caption{Average number of ones in the true transmission-response matrices for different pairs of parameters $(\Delta t_{1}, \Delta t_{2})$ in a simulation of spontaneous, in-vivo-like activity (Ca 1.2)}
\label{fig_1}
\end{figure*}

\begin{figure*}[!htp]
\centering
\includegraphics[scale=0.85,angle=0.00, clip=true]{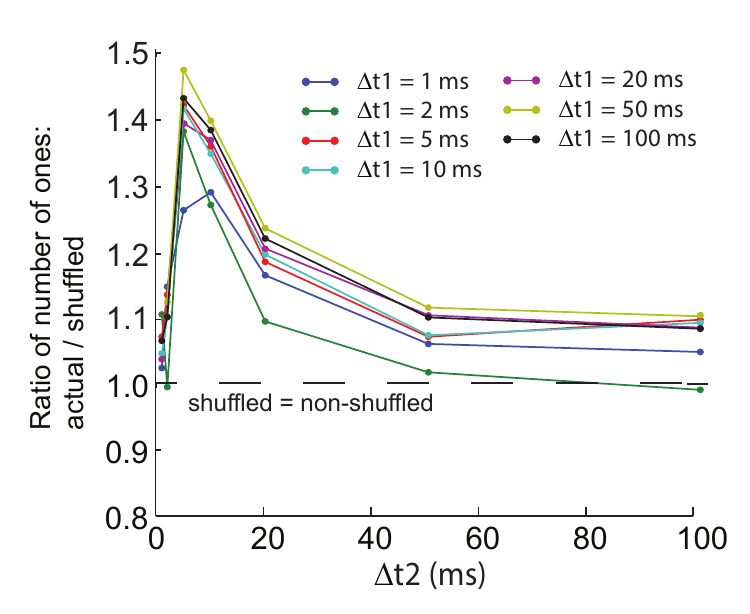}
\caption{Comparing randomized and non-randomized transmission-response matrices: average number of ones in a true transmission-response (t-r) matrix divided by the average number of ones obtained when the recorded spikes were randomized before calculating the t-r matrix. Matrices were calculated from simulated spontaneous, ongoing activity with different values for $\Delta t_{1}$ (in different colors) and $\Delta t_{2}$ (along the x-axis). For each pair $(\Delta t_{1}, \Delta t_{2})$,  matrices for 20 time steps were calculated, and the mean ratio is shown. Spikes were randomized by shuffling the identities of the firing neurons, thus conserving the number of spikes in any given time step and the total number of spikes fired by each neuron.}
\label{fig_2}
\end{figure*}

The average number of ones in the transmission-response matrix arising from simulated actitivity in the reconstructed microciruit, as a function of $\Delta t_{1}$ and $\Delta t_{2}$, is illustrated in Figure \ref{fig_1}.   Figure \ref{fig_2} shows the ratio of the average number of ones in the true transmission-response matrices  to the average number of ones in the simplified transmission-response matrices,  for various values of $\Delta t_{1}$ and $\Delta t_{2}$. In all cases we find that the maximum lies between $\Delta t_2 = 5 \text{ ms}$ and $\Delta t_2 = 10 \text{ ms}$, leading us to choose to work with $\Delta t_{2}=7.5 \text{ ms}$. For $\Delta t_1$ we find a maximum at 50 ms, but we use $\Delta t_{1}=5$ ms (for which the maximum ratio is only slightly lower than for $\Delta t_{1}=50$ ms) instead to avoid more than one spike per neuron per bin.

\section*{SM2. Gaussian Bayes classifiers}
Suppose there is a distribution $\rho$ over  $\mathbb{R} \times \{c_1,c_2,\dots,c_k\}$, where $ \{c_1,c_2,\dots,c_k\}$ is a set of class labels. We can project $\rho$ onto each of  the coordinates to construct a real-valued random variable $X$ and a class-label-valued random variable $Y$. We wish to build a classifier $C:\mathbb{R} \to \{ c_1,c_2,\dots,c_k \}$ which will, for any real number, choose the most likely class to which it might belong. That is, 
$$C(x)=\argmax_{c\in  \{c_1,c_2,\dots,c_k\} } P(Y=c|X=x),$$ 
where $P(A|B)$ is the probability of $A$ conditional on $B$ and $\argmax_{a\in A} f(a)$ denotes the element $a\in A$ such that $f(a)$ is maximal.  This element of $A$ will in practice always be unique.
% This means that the conditional distribution of $X$, given that the label $Y$ takes the value r is given by
% $X\mid Y=r \sim P_r for r=1,2,\dots,K$
% where ``$\sim$ '' means ``is distributed as'', and where P_r denotes a probability distribution.
% From this sample data we wish to build a classifier $C:\mathbb{R} \to Y$ which will, for any real number, chooses the most likely class for it to belong to.

% This is built using a data set consisting of samples from $(X,Y)$ as well as a prior distribution of the classes. 

% Given an instance $X=x$, the posterior probability over $\{c_1,c_2,\dots,c_k\}$ has are the conditional probabilities $P(Y=c|X=x)$.

Bayes' theorem states that 
$$P(Y=c|X=x)P(X=x)= P(X=x|Y=c)P(Y=c).$$ A Bayesian classifier picks the class with the highest conditional probability, which using Bayes' theorem is $$C(x)
% =\argmax_{c\in\{c_1, c_2, \ldots c_k\} }P(Y=c|X=x)
= \argmax_{c\in\{c_1, c_2, \ldots c_k\}} \frac{P(X=x|Y=c)P(Y=c)}{P(X=x)}.$$

Usually $\rho$ itself is unknown and must be infered from sample data. We then also assume some model distribution to estimate $\rho$ from these samples. The Gaussian Bayes classifier is the Bayes' classifier under the assumption that the distribution of each separate class is Gaussian.

After calculating the means and variances of the sample data within each of the classes separately, we model their respective distributions by the Gaussians $N(\mu_{c_i}, \sigma_{c_i}^2)$. If $p(A)$ denotes the probability density function of $A$, then
\begin{align*}
\frac{P(X=x|Y=c)}P(Y=c){P(X=x)}&=\frac{p(X=x|Y=c)}P(Y=c){p(X=x)}\\
&=\frac{1}{\sqrt{2\pi \sigma_c^2}}\exp \left( \frac{-(x-\mu_c)^2} {2\sigma_c^2}\right)\frac{P(Y=c)}{p(X=x)}
\end{align*}

% Note that the factor is $1/p(X=x)$ common to all classes and thus does not affect which class achieves the maximum.

A common situation, such as in our analysis, is a uniform prior. A uniform prior over $\{c_1,c_2,\dots,c_k\}$ means $P(Y=c_i)=1/k$ for all $i$. If we assume a uniform prior, then the factor $\frac{P(Y=c)}{p(X=x)}$ is common to all classes and thus does not affect which class achieves the maximum. Thus we get the formula
\begin{align*}
C(x) 
% &=\argmax_{c\in\{c_1, c_2, \ldots c_k\} }P(Y=c|X=x)\\
% &=\frac{P(X=x|Y=c)P(Y=c)}{P(x)}\\
% &=\argmax_{c\in\{c_1, c_2, \ldots c_k\}} \frac{p(X=x|Y=c)P(Y=c)}{p(X=x)} \\
&=\argmax_{c\in\{c_1, c_2, \ldots c_k\}} \frac{1}{\sqrt{2\pi\sigma_c^2}}
\exp\left(\frac{-(x-\mu_c)^2}{2\sigma_c^2} \right).
\end{align*}

% Bayes' rule says that, for any test data $x$ and class $c$,
% $$P(c|x)\propto  􏰄\frac{1}{\sqrt{2\pi \hat{\sigma}_c^2}}
% \exp􏰂\left(\frac{-(x-\hat{\mu}_c)^2}{2\hat{\sigma}_c^2} \right)
% 􏰃P(c).$$
% Assuming a uniform prior and Gaussian distributions, Bayes' theorem provides a classifier
% $$\text{Class}(x)=
% \argmax_{c\in \{\text{dot}, \text{circle}\}}􏰄
% \frac{1}{\sqrt{2\pi\hat{\sigma}_c^2}}
% \exp􏰂\left(\frac{-(x-\hat{\mu}_c)^2}{2\hat{\sigma}_c^2} \right) .$$
% With a Monte-Carlo simulation we calculated the expected percentage of correct classifications under different random splittings of the data into training and testing sets.

\vfill\eject

\section*{SM3. Randomization of connection matrices and other control cases}
We created four types of random matrices of sizes and connection probabilities similar to the connectivity matrices of the BBP reconstruction. 

\subsection*{SM3.1. Generation of  Erd\H os-R\'enyi random matrices}
For this basic control we first computed the overall connection probability in the reconstruction and found it to be $0.8\%$. We then generated random, binary square matrices of size $3.1\times  10^4$, where 1's were placed at random off-diagonal in the matrix  with  probability $0.8\%$.

\subsection*{SM3.2. Randomization preserving the distance-dependent connectivity between layers}
Input for this randomization method were the structural matrix and the matrix of pairwise soma distances, both generated as part of the standard BBP reconstruction process.
The rows and columns of both matrices were first grouped into $N=6$ groups according to the layer of the neuron they correspond to. This effectively partitioned both matrices into $N*N=36$ submatrices each. For each pair of submatrices, the soma distances were grouped into  bins of size $25 \mu m$. Next, in the submatrix corresponding to each distance bin, we first replaced all 1's by 0's  and then replaced randomly chosen 0's by 1's, so that the total number of 1's was preserved. Creation of autapses, i.e., a connection from a neuron to itself, was avoided by creating a separate bin for distances of $0 \mu m$.

The result was a connection matrix with the same number of connections between each pair of layers  and the same distance-dependent connection probability between pairs of layers, to within $25 \mu m$, as the original matrix.

\subsection*{SM3.3. Randomization preserving the distance-dependent connectivity between m-types}
This randomization method was identical to the preceding randomization, preserving connectivity between layers, except that  the neurons were partitioned initially into $N=55$ groups of morphological types instead of only six layers.

\subsection*{SM3.4. Generation of  connection matrices according to Peters' Rule}
For this control case, we started with a connection matrix that placed a connection not just where a synaptic connection was found in the reconstructed microcircuit, but between each pair of neurons whose arbors came within close proximity (closer than $3 \mu m$). The resulting matrix had approximately 16 times more connections than the structural matrix. These connections were then pruned randomly with a uniform probability until the same number of connections as in the structural matrix was attained.

\vfill\eject

\section* {Supplementary Text}

\begin{center}
{\textbf {Supplementary Text}}\\
To accompany ``Topological analysis of the connectome of digital reconstructions of neural microcircuits."\\
Pawe\ l\ D\l otko, Kathryn Hess, Ran Levi, Max Nolte, Michael Reimann, Martina Scolamiero, Katharine Turner, Eilif Muller, Henry Markram
\end{center}
\bigskip

\section*{ST1. The topological toolbox}

Most of the mathematical methods we describe here are part of the  basic toolbox of algebraic topology, though perhaps not as well known in the directed variants presented here. We give a brief account of these concepts for the benefit of the non-expert, and refer to literature for the reader interested in further details. 

We explain first how to associate to any directed graph a simplicial complex known as its \hadgesh{directed flag complex}, then recall two types of important invariants of simplicial complexes, which turn out to be very useful for analyzing the digitally reconstructed microcircuits: the Euler characteristic and Betti numbers.   We then describe the data structures and algorithms that we implemented in order to construct the flag complexes of the directed graphs representing the microcircuits and to compute their Euler characteristics and Betti numbers.  

\subsection*{ST1.1. Directed graphs}

 A \hadgesh{directed graph} $\calg$ consists of a pair of finite sets $(V,E)$ and a function $\tau\colon E\to V\times V$. The elements of the set $V$ are the \hadgesh{vertices} of $\calg$, the elements of $E$ are the \hadgesh{edges} of $\calg$, and the function $\tau$ associates with each edge an ordered pair of vertices. The \hadgesh{direction} of an edge $e$ with $\tau(e) = (v_1,v_2)$ is taken to be from $\tau_{1}(e)=v_1$, the \hadgesh{source vertex}, to $\tau_{2}(v)=v_2$, the \hadgesh{target vertex}. The function $\tau$ is required to satisfy the following two conditions.
 \begin{enumerate}
 \item For each $e\in E$, if  $\tau(e) = (v_1, v_2)$, then $v_1\neq v_2$, i.e., there are no loops in the graph.
  \item The function $\tau$ is injective, i.e., for any pair of vertices $(v_{1},v_{2})$, there is at most one edge directed from $v_{1}$ to $v_{2}$. 
 \end{enumerate}
  A vertex $v\in\calg$ is said to be a \hadgesh{sink} if $\tau_{1}(e) \neq v$ for all $e\in E$,  and a \hadgesh{source} is if $\tau_{2}(e)\neq v$ for all $e\in E$.  
 
\begin{figure*}[!ht]
  \centering
      \includegraphics[width=0.75\textwidth]{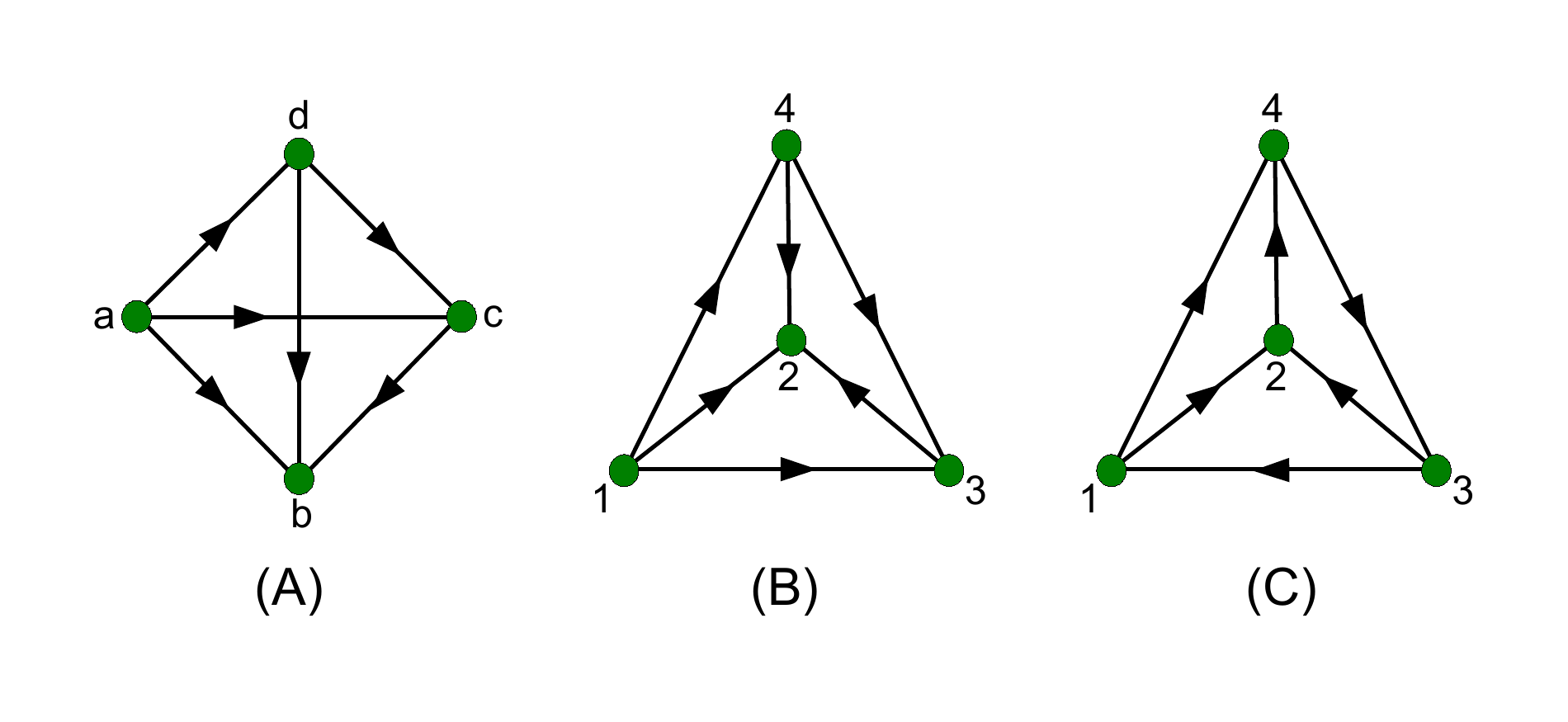}
  \caption{(A-C) Examples of directed graphs. Graphs (A) and (B) are isomorphic, where the isomorphism is given by the map sending vertex $a$ to 1, $b$ to 2, $c$ to 3, and $d$ to 4. Graphs (A) and (B) are not isomorphic to graph (C). Vertex $b$ in graph (A) is a sink, vertex $a$ in the same graph is a source. Graph (C) has no sources or sinks, which explains the lack of isomorphism to graphs (A) and (B).}\label{fig:graphs}
\end{figure*}

 To compare two graphs, we require the following notion.  A \hadgesh{morphism of directed graphs} from a directed graph $\calg=(V,E, \tau)$ to a directed graph $\calg'=(V', E', \tau ')$ consists of a pair of set maps $\alpha:V\to V'$ and $\beta: E \to E'$ such that $\beta$ takes an edge in $\calg$  with source $v_{1}$ and target $v_{2}$ to an edge in $\calg'$ with source $\alpha(v_{1})$ and target $\alpha(v_{2})$, i.e.,  $\tau'\circ \beta = (\alpha, \alpha)\circ\tau$.  Two graphs  $\calg$ and $\calg'$ are \hadgesh{isomorphic} if there is morphism of graphs $(\alpha, \beta): \calg \to \calg'$ such that both $\alpha$ and $\beta$ are bijections, which we call an \hadgesh{isomorphism of directed graphs} (Figure \ref{fig:graphs}).

 A \hadgesh{path} in a directed graph $\calg$ consists of a sequence of edges $(e_{1},..., e_{n})$ such that  for all $1\leq k< n$, the target of $e_{k}$ is the source of $e_{k+1}$, i.e., $\tau_{2}(e_{k})=\tau_{1}(e_{k+1})$. The \hadgesh{length} of the path $(e_{1},..., e_{n})$ is $n$, the number of edges of which the path is composed.   If, in addition, target of $e_{n}$ is the source of $e_{1}$, i.e., $\tau_{2}(e_{n})=\tau_{1}(e_{1})$, then $(e_{1},..., e_{n})$ is an   \hadgesh{oriented cycle}.

\subsection*{ST1.2. Simplicial complexes}\label{sec:simplicial}
An \hadgesh{abstract oriented simplicial complex} is a  collection $\cals$ of finite, {ordered} sets with the property that if $\sigma \in \mathcal{S}$, then every subset $\tau$ of $\sigma$ is also a member of $\mathcal{S}$. A \hadgesh{subcomplex} of an abstract oriented simplicial complex is a sub-collection $\cals'\subseteq\cals$ that is itself an abstract oriented simplicial complex.  
Henceforth, we simplify terminology and usually refer to abstract oriented simplicial complexes merely as simplicial complexes. 

The elements of a simplicial complex $\mathcal{S}$ are called its \hadgesh{simplices}.  A simplicial complex is said to be \hadgesh{finite} if it has only finitely many simplices. If $\sigma \in \mathcal{S}$, we define  the \hadgesh{dimension} of $\sigma$, denoted $\dim(\sigma)$, to be $|\sigma|-1$, the cardinality of the set $\sigma$ minus one. If $\sigma$ is a simplex of dimension $n$, then we refer to $\sigma$ as an \hadgesh{$n$-simplex} of $\cals$. The set of all $n$-simplices of $\mathcal{S}$ is denoted $\mathcal{S}_n$. A simplex $\tau$ is said to be a \hadgesh{face} of $\sigma$ if $\tau$ is a subset of $\sigma$ of a strictly smaller cardinality.   A \hadgesh{front face} of an $n$-simplex $\sigma=(v_{0},...,v_{n})$ is a face $\tau=(v_{0},...,v_{m})$ for some $m<n$. Similarly, a \hadgesh{back face} of $\sigma$ is a face $\tau' = (v_{i},\ldots, v_n)$ for some $0<i<n$. If $\sigma= (v_0, \ldots, v_n)\in \mathcal S_{n}$, then the \hadgesh{$i^{\text{th}}$ face} of $\sigma$ is the $(n-1)$-simplex $\sigma^i$ obtained from $\sigma$ by removing the vertex $v_i$. 

A simplicial complex gives rise to a topological space by means of the construction known as \hadgesh{geometric realization}. In brief, one associates a point (a standard geometric 0-simplex) with each 0-simplex,  a line segment (a standard geometric 1-simplex) with each 1-simplex, a filled-in triangle (a standard geometric 2-simplex) with each 2-simplex, etc., glued together along common faces. The intersection of two simplices in $\cals$, neither of which is a face of the other, is a proper subset, and hence a face, of both of them. In the geometric realization this means that the geometric simplices that realize the abstract simplices intersect on common faces, and hence give rise to a well-defined geometric object. A geometric $n$-simplex is nothing but a   $(n+1)$-clique, canonically realized as a geometric object. An $n$-simplex is said to be \emph{oriented} if there is a linear ordering on its vertices. In this case the corresponding $(n+1)$-clique is said to be a \emph{directed $(n+1)$-clique}.

If $\cals$ is a simplicial complex, then the union $\cals^{(n)} = \cals_n\cup\cdots \cup \cals_0$, which is called the \hadgesh{$n$-skeleton} of $\cals$, is a subcomplex of $\cals$. We say that $\cals$ is \hadgesh{$n$-dimensional} if $\cals = \cals^{(n)}$, and $n$ is minimal with this property. 
If $\cals$ is $n$-dimensional, and $k\le n$, then the collection $\cals_k\cup\ldots\cup\cals_n$ is not a subcomplex of $\cals$ because it is not closed under taking subsets. However if one adds to that collection all the faces of all simplices in $\cals_k\cup\ldots\cup\cals_n$, one obtains a subcomplex of $\cals$ called the \hadgesh{$k$-coskeleton} of $\cals$, which we will denote by $\cals_{(k)}$. The computational usefulness of coskeleta will become clear when we discuss homology computation  (ST1.3).

Directed graphs give rise to abstract oriented simplicial complexes in a natural way. Let $\calg = (V, E, \tau)$ be a directed graph. The \hadgesh{directed flag complex} associated to $\calg$ is the abstract simplicial complex $\cals = \cals(\calg)$, with $\cals_0=V$  and whose 
$n$-simplices $\cals_n$ for $n\geq 1$ are $(n+1)$-tuples $(v_0,\ldots,v_n)$, of vertices such that for each $0\le i< j \le n$, there is an edge in $\calg$ from $v_i$ to $v_j$. Notice that because of the assumptions on $\tau$, an $n$-simplex in $\cals$ is characterised by the (ordered) sequence $(v_0,\ldots,v_n)$, but not by the underlying set of vertices. For instance $(v_1, v_2, v_3)$ and  $(v_2,v_1,v_3)$ are distinct $2$-simplices with the same set of vertices.

\subsection*{ST1.3. Homology, Betti numbers, and Euler characteristic}
We now recall certain well known invariants of simplicial complexes arising in algebraic topology, which are preserved under  a class of morphisms that is relevant in algebraic topology and that includes isomorphisms.  These invariants serve to measure the ``complexity'' of simplicial complexes, from various topological perspectives, leading us to refer to them as \emph{metrics}.  

Homology is an important algebraic invariant of topological spaces. In this paper we use only \hadgesh{mod-2 simplicial homology}, computationally the simplest variant of homology, which is why we choose to work with it in applications, though other types of simplicial homology may provide deeper information. We do not give a complete account of homology here, but rather an elementary description of what it is and its basic properties. 

Let $\F_2$ denote the field of two elements, which we denote by $0$ and $1$. Let $\cals$ be a finite simplicial complex. Define the \hadgesh{chain complex} $C_*(\cals,\F_2)$ to be the sequence  $\{C_n = C_n(\cals, \F_2)\}_{n\geq 0}$, such that $C_n$ is the $\F_{2}$-vector space whose basis elements are the $n$-simplices $\sigma\in\cals_n$, for each $n\geq 0$. In other words, the elements of $C_n$ are formal linear combinations of $n$-simplices in $\cals$ with coefficients in $\F_2$. For each $n\geq 0$, there is a linear transformation called a \hadgesh{differential} 
\[\partial_n\colon C_{n+1}\to C_n\]
defined by $\partial_n(\sigma) = \sigma^0 + \sigma ^1 +\cdots + \sigma^n$ for every $n$-simplex $\sigma$, where $\sigma^i$ is the $i$-th face of $\sigma$, as defined above. Having defined $\partial_n$ on the basis, one extends the definition linearly to the entire vector space $C_n$.

The $n$-th Betti number $\beta_n(\cals)$ of a simplicial complex $\cals$ is  the $\F_2$-vector space dimension of its $n$-th mod 2 \hadgesh{homology group}, which is defined by 
\[H_n(\cals, \F_2) = \Ker(\partial_{n-1})/ \mathrm{Im}(\partial_n).\]

Computing the Betti numbers is conceptually very easy. Let $|\cals_{n}|$ denote the number of $n$-simplices in the simplicial complex $\cals$. If one encodes the differential $\partial_n$ as a $\big(|\cals_{n}|\times |\cals_{n+1}|\big)$-matrix $D_n$ with coefficients in $\F_{2}$, then one can easily compute its \hadgesh{nullity}, $\nll(\partial_n)$, and its \hadgesh{rank}, $\rk(\partial_n)$, which are the $\F_2$-dimensions of the null-space and the column space of $D_n$, respectively. The  \hadgesh{Betti numbers} of $\cals$ are then a sequence of natural numbers defined by 
\[\beta_0(\cals) = \dim_{\F_2}(C_0) - \rk(\partial_0),\quad \text{and} \quad \beta_n(\cals) = \nll(\partial_{n-1}) -\rk(\partial_{n}).\]

The $n$-the Betti number $\beta_n$ counts the number of ``$n$-dimensional holes'' in the geometric realization of $\cals$.  When $\cals =\cals(\calg)$ is the directed flag complex of a directed graph $\calg$,  both the simplices of $\cals$ and these ``$n$-dimensional holes'' can be regarded as particularly important ``metamotifs'' \cite{motif} in the graph $\calg$. 

It is easy to show that the $n$-th Betti number of a simplicial complex $\cals$ is equal to that of its $(n-1)$-st coskeleton $\cals_{(n-1)}$, i.e.,  $\beta_{n}(\cals)=\beta_{n}(\cals_{(n-1)})$, for all $n$.  This observation turns out to be computationally very useful, since there is no need to store the simplices of dimension less than $n-1$ that are not faces of higher dimensional simplices in order to compute $\beta_{n}(\cals)$. In this paper it was exactly this trick that allowed us to compute the top dimensional homology of the 42 N-complexes we worked with.

Homology actually encodes far more information than what is intimated here, which can potentially be used for analyzing networks, but for the purposes of this article the description above will suffice.

If $\cals$ is a simplicial complex and $|\cals_n|$ denotes the cardinality of the set of $n$-simplices in $\cals$, then the Euler characteristic of $\cals$ is defined to be 
\[\chi(\cals) = \sum_{n\geq 0} (-1)^n|\cals_n|.\]
There is a well known, close relationship between Euler characterstic and Betti numbers \cite{hatcher}, which is expressed as follows.
If $\{\beta_n\}_{n\geq 0}$ is the sequence of Betti numbers for $\cals$, then 
\[\chi(\cals) = \sum_{n\geq 0} (-1)^n\beta_n(\cals).\]
See Figure \ref{fig:figure2}A for a specific example.

\subsection*{ST1.4. Hasse Diagrams}
A \hadgesh{Hasse diagram}, otherwise known as a {directed acyclic graph}, is a directed graph $\calh=(V,E,\tau)$ with no oriented cycles. Hasse diagrams can be used to encode various combinatorial, geometric, and topological structures, such as posets and cubical complexes. Below we explain in detail how Hasse diagrams encode simplicial complexes. We include this discussion here because our computational algorithm (Algorithm 1) is based on this idea. 

A Hasse diagram $\calh$ is said to be \hadgesh{stratified} if for each $v\in V$, every path from $v$ to any sink has the same length. Thus in a stratified Hasse diagram the vertices are naturally partitioned into disjoint strata, where every directed path from a vertex in the $k$-th stratum $V_{k}$ to any sink is of length $k$. In particular, the $0$-th stratrum $V_{0}$ is the set of sinks of $\calh$.    Moreover, for all $e\in E$, there exists $k>0$ such that $\tau_{1}(e)\in V_{k}$ and $\tau_{2}(e)\in V_{k-1}$.  Note that if $\calh$ and $\calh'$ are isomorphic Hasse diagrams, and $\calh$ is stratified, then so is $\calh'$.

An \hadgesh{orientation} $\varsigma$ on a Hasse diagram $\calh$ consists of a linear ordering $<_{\varsigma, v}$ of the set $E_{v}$ of edges with source $v$, for every vertex $v$ of $\calh$.  If $\calh=(V,E,\tau)$ and $\calh'=(V',E', \tau')$ are Hasse digrams equipped with orientations $\varsigma$ and $\varsigma'$, respectively, then a  \hadgesh{morphism of oriented Hasse diagrams} from $(\calh , \varsigma)$ to $(\calh', \varsigma')$ is a morphism of directed graphs $(\alpha, \beta): \calh \to \calh'$ such that for every $v\in V$, the restriction of $\beta$ to a set map $E_{v}\to E_{\alpha(v)}$ preserves the orientation, i.e, if $e<_{\varsigma, v} e'$ for some $e,e'\in E_{v}$, then $\beta (e) <_{\varsigma', \alpha(v)} \beta (e')$.  A morphism $(\alpha, \beta)$ of oriented Hasse diagrams is an \hadgesh{isomorphism} if $\alpha$ and $\beta$ are bijections. A stratified Hasse diagram equipped with an orientation is called \hadgesh{admissible}.

\begin{figure*}[!ht]
  \centering
      \includegraphics[width=0.5\textwidth]{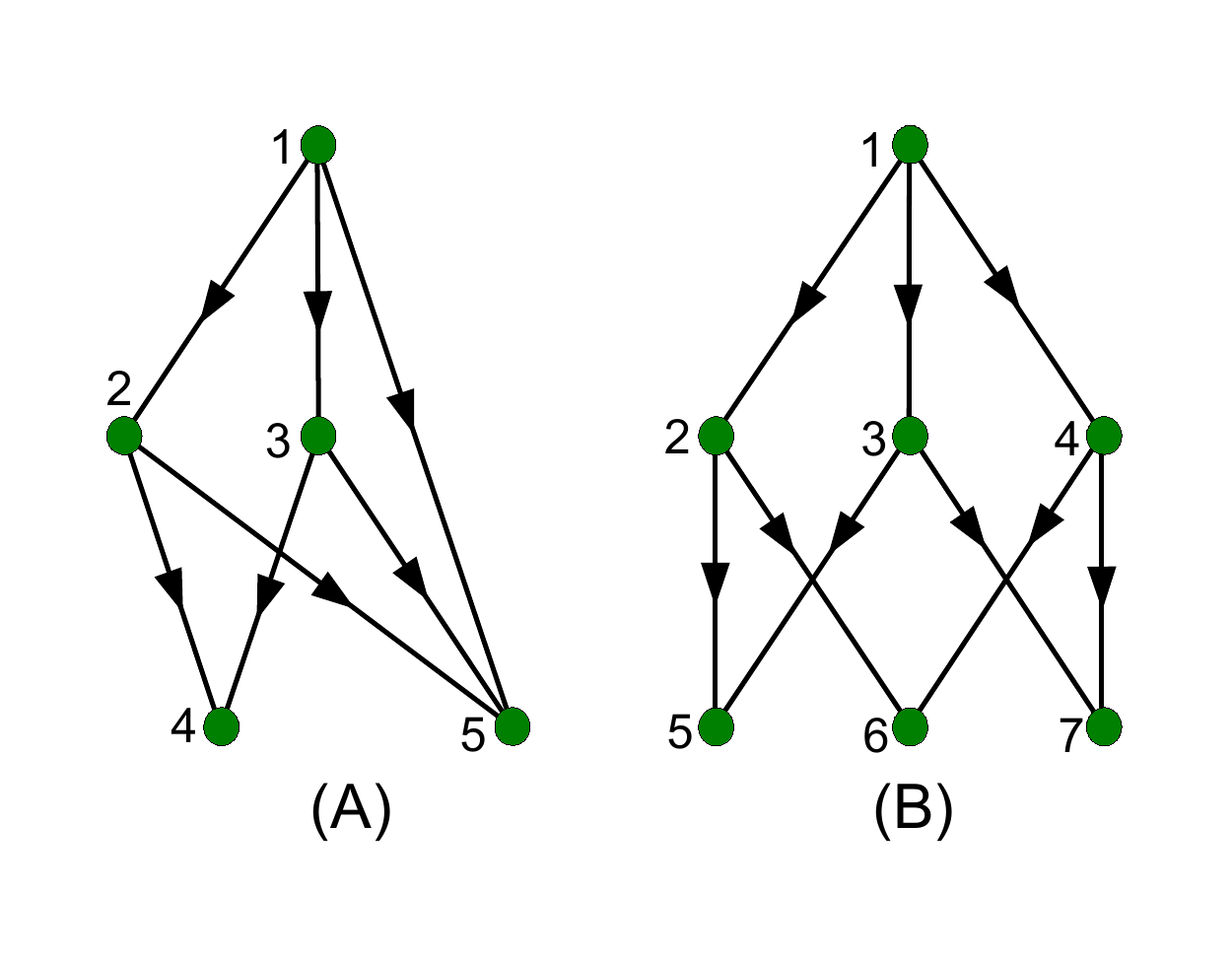}
  \caption{(A)	A Hasse diagram that is not stratified, due to the edge from the vertex 1 to 5. (B) A stratified Hasse diagram, where vertices 5, 6, and 7 are the vertices of level 0, vertices 2, 3, and 4 are of level 1, and vertex 1 is of level 2. This is also an admissible Hasse diagram, where the outgoing edges are ordered from left to right. Vertex 2  is a front face of vertex 1, while vertex 3 is neither a front nor a back face of a vertex 1, and vertex 4 is back face of a vertex 1.}\label{fig:hasse}
\end{figure*}

Vertices in the $k$-th stratum of a stratified Hasse diagram $\calh$ are said to be \hadgesh{of level $k$}. If $k<n$, and $v, u$  are vertices of levels $k$ and $n$ respectively, then we say that $v$ is a \hadgesh{face} of $u$ if there is a path in $\calh$ from $u$ to $v$.  If $\calh$ is also oriented and therefore admissible, and there is a path $(e_{1},...,e_{n-k})$ from $u$ to $v$ such that $e_{i}=\min E_{\tau_{1}(e_{i})}$ for all $1\leq i\leq n-k$, we say that $v$ is a \hadgesh{front face} of $u$.  Similarly, $v$ is a \hadgesh{back face} of $u$ if there is a path $(e_{1},...,e_{n-k})$ from $u$ to $v$ such that $e_{i}=\max E_{\tau_{1}(e_{i})}$ for all $1\leq i\leq n-k$.   We let $\operatorname{Face}(u)$ denote the set of all faces of $u$ and $\operatorname{Face}(v)_{k}$ the set of those that are of level $k$, while $\operatorname{Front}(u)$ and $\operatorname{Back}(u)$ denote its sets of front and back faces, respectively.  See Figure \ref{fig:hasse} for an illustration of the concepts introduced above.

\begin{Ex}\label{every_graph_is_a_shd}
If  $\calg = (V,E, \tau)$ is a directed graph, then $\calg$ can be equivalently represented by an admissible Hasse diagram with level 0 vertices $V$, level 1 vertices $E$, and directed edges from each $e\in E$ to its source and target. The ordering on the edges in the Hasse diagram is determined by the orientation of each edge $e$ in $\calg$.
\end{Ex}

\begin{figure*}[!ht]
  \centering
      \includegraphics[width=0.75\textwidth]{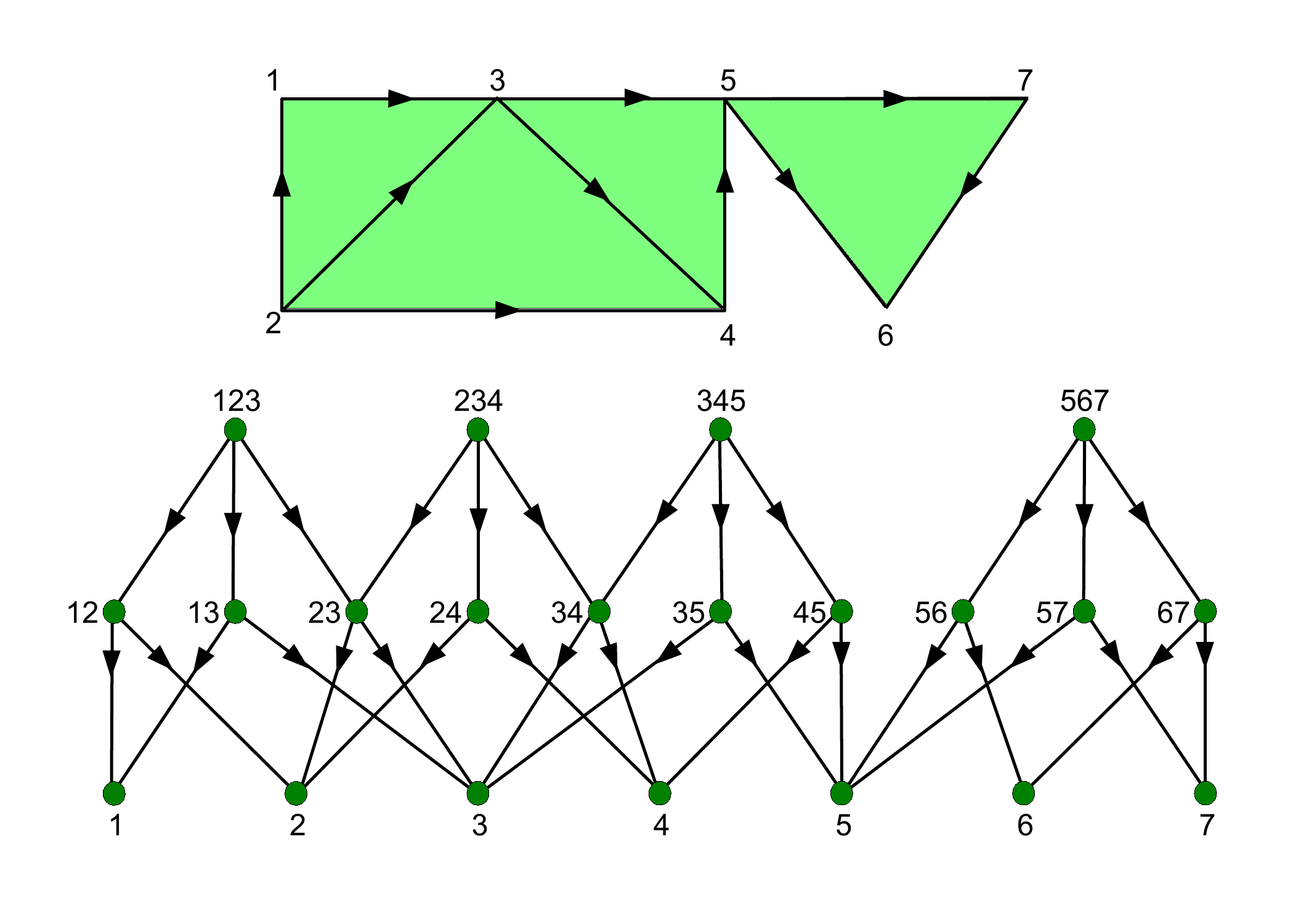}
  \caption{Top: The geometric realization of a  simplicial complex consisting of seven 0-simplices (labeled 1,...,7), ten 1-simplices, and four 2-simplices. The orientation on the edges is denoted by arrows, i.e., the tail of an arrow is its source vertex, while the head of an arrow is its target. 
Bottom: The Hasse diagram corresponding to the simplicial complex above. Level $k$ vertices correspond to $k$-simplices of the complex and are labeled by the ordered sets of vertices that constitute the corresponding simplex. Note that, e.g.,  vertex 23 is a back face of a vertex 123 and a front face of a vertex 234. 
}\label{fig:hasse-sc}
\end{figure*}

Every simplicial complex $\cals$ gives rise to an admissible Hasse diagram $\calh_\cals$ as follows. The level $d$ vertices of $\calh_\cals$ are the $d$-simplices of $\cals$. There is a directed edge from each $d$-simplex to each of its $(d-1)$-faces. The stratification on $\calh_\cals$ is thus given by dimension, and the orientation is given by the natural ordering of the faces of a simplex from front to back.  See Figure \ref{fig:hasse-sc}.

The \hadgesh{Euler characteristic} of a stratified Hasse diagram $\calh=(V, E, \tau)$ is defined to be the integer
\[\chi(\calh) = \sum_{k\geq 0} (-1)^k |V_k|.\]
It is easy to see that isomorphic stratified Hasse diagrams have the same Euler characteristic. It is also straight forward to show that if $\calh$ is a stratified Hasse diagram associated to a simplicial complex $\cals$, then the Euler characteristic of $\calh$ coincides with that of $\cals$.

\section*{ST2. Data structures and algorithms}
In this section we describe our basic data structures and provide a detailed overview of the algorithm that constructs the directed flag complex associated to a directed graph.  We also indicate briefly how our homology computations were performed.  A publicly available C++ implementation of the code will be available on http://neurotop.gforge.inria.fr/.

\newcommand{\Ver}{\mathrm{Ver}}
\newcommand{\Tar}{\mathrm{Tar}}
\newcommand{\Src}{\mathrm{Src}}
\newcommand{\lev}{\mathrm{L}}
\newcommand{\Wgt}{\mathrm{Wgt}}
\newcommand{\intgr}{\mathrm{int}}
\newcommand{\dbl}{\mathrm{double}}

\subsection*{ST2.1. Data structures}
\label{sec:dataStructures}
We represent an admissible Hasse diagram $\calh$ corresponding to the directed flag complex of a directed graph $\calg=(V,E, \tau)$ by a reference-based data structure, using vectors to store the references to the vertices of the diagram. Each vertex $v\in \calh$ stores the following information.
\begin{enumerate}
\item $\Ver(v)$: A vector of the vertices of $\calg$ determining the simplex of the flag complex to which $v$ corresponds.
\item $\Tar(v)$: A vector of references to the vertices that are targets of edges with source $v$.
\item $\Src(v)$: A vector of references to the vertices that are sources of edges with target $v$.
%\item $\lev(v)$: The level of $v$. 
\end{enumerate}
The admissible Hasse diagram $\calh$ is thus represented by an ordered set of $d$ vectors, where $d$ is the maximal level in $\calh$, and where the $i$-th vector contains the references to all level $i$ vertices.  

Let $S_{\intgr}$ denote the size of integer data types, and for a given graph $\calg = (V,E,\tau )$, let $|V|$ and $|E|$ denote the cardinalities of the corresponding sets. Each edge of the Hasse diagram is stored in two vertices of the diagram. If each reference requires $S_{\intgr}$ storage, then  we require $O(|E|\cdot S_{\intgr})$ space to store all references. In addition, each vertex stores 
%its level and 
the vector of vertices in $V$ of the simplex in the flag complex of $\calg$ to which it corresponds, which requires an additional $O( S_{int}\cdot d )$ of space per vertex. The total size of a Hasse diagram is thus bounded by $O( ( S_{\intgr}\cdot d)\cdot |V| + |E|\cdot S_{\intgr} )$.  In particular, the required storage space grows linearly with the number of vertices and with the number of edges. For our complexity analysis below we assume that accessing any vertex, using $\Tar$ or $\Src$, takes $O(1)$ time.

\begin{algorithm}[h!]
  \small
  \caption{Directed flag complex generation.}
  \label{alg:generateDirectedFlagComplex}
  \begin{algorithmic}[1]
	\REQUIRE A directed graph $\calg=(V,E,\tau)$.
	\ENSURE A Hasse diagram $\calh$ representing the directed flag complex associated to $\calg$.
	\STATE Convert $\calg$ to level 0 and level 1 vertices of $\calh$ (cf.~Example \ref{every_graph_is_a_shd}). \label{start}
	\FOR { every level 1 vertex $e\in\calh$ }\label{alg:generateDirectedFlagComplex:firstForLoop}
		\IF { exist $e_1$, $e_2$ such that $\tau_1(e_1) = \tau_1(e)$, $\tau_1(e_2) = \tau_2(e)$ and $\tau_2(e_1) = \tau_2(e_2) = u$  }\label{alg:generateDirectedFlagComplex:IfStatementInTheForLoop}
			\STATE Add $u$ to $U_e$;
		\ENDIF
	\ENDFOR
	
	\STATE $dim = 2$;
	\REPEAT \label{alg:generateDirectedFlagComplex:repeatUntilLoop}
		\STATE \emph{next\_level\_nodes} -- empty vector of references to nodes;
		\FOR {top--level vertex $e\in \calh$ }
			\FOR { Every $u \in U_e$ }
				\STATE Create a node $t$ of a Hasse diagram;
				\STATE $\Ver(t) = [\Ver(e),u]$; \label{alg:generateDirectedFlagComplex:extendingOrderingForLargerSimplex}
				\STATE $U_t = U_e$;
				%\STATE $\lev(t) = dim$; 
				\STATE Add $e$ to $\Tar(t)$;
				\STATE Add $t$ to $\Src(e)$;
				\FOR { Every $bd \in \Tar(e)$ }\label{alg:generateDirectedFlagComplex:targetSourceSearch}
					\FOR { Every $cbd \in \Src(bd)$ }
						\IF { $u \in \Ver(cbd)$ }
							\STATE Add $cbd$ to $\Tar(t)$;
							\STATE Add $t$ to $\Src(cbd)$;
							\STATE $U_t = U_t \cap U_{cbd}$;
						\ENDIF
					\ENDFOR
				\ENDFOR
				\STATE Add $t$ to \emph{next\_level\_nodes};
			\ENDFOR
		\ENDFOR
	    \STATE Add \emph{next\_level\_nodes} to $\calh$;
	    \STATE $dim = dim + 1$;
	\UNTIL{\emph{next\_level\_nodes} = $\emptyset$}
	\STATE Return $\calh$;
  \end{algorithmic}
\end{algorithm}

\subsection*{ST2.2. Creation of the directed flag complex associated to a directed graph}
We describe our algorithm that creates a directed simplicial complex given a directed graph $\calg$.  The output  is a Hasse diagram $\calh$, stored as the data structure described above. The identifier $\Ver(v)$ of a vertex $v$ in $\calh$, corresponding to a simplex $\sigma$ in the directed flag complex, is the vector of vertices in $\calg$ that represents $\sigma$. 

For every level $n\geq 1$ vertex $v$ in $\calh$ such that $\Ver(v) = [v_0,\ldots,v_n]$, the algorithm additionally records a vector $U_v$ of references to  level $0$ vertices $u$ satisfying the following properties:
\begin{enumerate}
\item $u \not = v_{i}$ for all $0\leq i\leq n$, and
\item for every $u \in U_v$ and  every $0\leq i\leq n$, there exists an edge in $\calg$ from $v_i$ to $u$. 
\end{enumerate}

Finally, we assume that the graph $\calg$ itself is given as an admissible Hasse diagram, as described in Example \ref{every_graph_is_a_shd}.
Under these assumptions Algorithm~\ref{alg:generateDirectedFlagComplex} below is used to create the directed flag complex associated to $\calg$.

\subsection*{ST2.3. Discussion of Algorithm~\ref{alg:generateDirectedFlagComplex}} 
At the start of the algorithm (Line~\ref{start}) only levels 0 and 1 of the Hasse diagram $\calh$, which are the same as those of the Hasse diagram representation of $\calg$ itself, have been created (cf.~Example \ref{every_graph_is_a_shd}). The  \emph{for} loop in the line~\ref{alg:generateDirectedFlagComplex:firstForLoop} initialises the creation of the vectors $U_v$ for level 1 vertices. For every level 1 vertex $e$, the vector $U_e$ stores the references to all the $0$-simplices that, together with $e$, will form a level $2$ vertex $t$. The construction of level $2$ vertices in $\calh$ is performed during the first iteration of the \emph{repeat-until} loop starting in Line~\ref{alg:generateDirectedFlagComplex:repeatUntilLoop}. 

We analyze the generation of level 2 vertices as a generic case, since the arguments may clearly be generalised to higher levels. The \emph{if} condition in  Line~\ref{alg:generateDirectedFlagComplex:IfStatementInTheForLoop} ensures that the vertex $u$ will be the terminal vertex of the $2$-dimensional simplex corresponding to the level 2 vertex $t$, created in the first iteration of the \emph{repeat-until} loop (Line~\ref{alg:generateDirectedFlagComplex:repeatUntilLoop}). Moreover the level $1$ vertex $e$ will correspond to a front face of the 2-simplex associated to $t$. Therefore, the ordering of $\Ver(e)$ can be extended to ordering of $\Ver(t)$, as in Line~\ref{alg:generateDirectedFlagComplex:extendingOrderingForLargerSimplex}.  Thus all level 2 vertices corresponding to 2-simplices in the directed flag complex of $\calg$ will be created by the algorithm. Also, since every simplex has a unique 1-dimensional front face, every $2$-simplex will be created only once by this process. 
 
Notice also that the \emph{if} condition in Line~\ref{alg:generateDirectedFlagComplex:IfStatementInTheForLoop} ensures that only triangles in $\calg$ consisting of three edges oriented as $(v_{1}, v_{2})$, $(v_{2},v_{3})$, and $(v_{1}, v_{3})$ will give rise to level 2 vertices in $\calh$. It follows by induction that the analogous condition on orientations is then automatically satisfied for simplices of dimension greater than $2$. To see this, fix $n\geq 2$, and suppose that all simplices of dimension less than or equal to $n$ have the desired property. Fix an $n$-simplex $S = [v_0,\ldots,v_n]$ and $u \in U_S$.  By definition of the set $U_S$, there is an edge from $v_i$ to $u$ for every $i \in \{0,\ldots,n\}$. Note that $u \in U_{S'}$ for  any $S' \in \Tar(S)$. The previous iteration of the \emph{repeat-until} loop (Line~\ref{alg:generateDirectedFlagComplex:repeatUntilLoop}) created an oriented simplex from $S'$ together with $u$, of which $u$ is the last vertex. Since the ordering of elements in $S'$ is a restriction of the ordering of elements in $S$,  the ordering of a $n+1$ dimensional simplex $[v_0,\ldots,v_n,u]$ restricted to any face yields the orientation of that face. It follows that Algorithm~\ref{alg:generateDirectedFlagComplex} does indeed construct a directed flag complex.
%Let us take a simplex $S'$ of such a type, that do not contain $v_0$. We know that $S \cap S'$ is properly oriented and that there are out-going edges in $\calg$ from $v_0$ to every vertex of $S \cap S'$. Moreover, we know that from every vertex of $S \cap S'$ there are outgoing edges to $u$. Therefore, there exist a unique ordering of a simplex $[v_0,\ldots,v_n,u]$ that restrict to the proper ordering of its subsimplices.  

We now discuss the termination of Algorithm~\ref{alg:generateDirectedFlagComplex}. If a level $n$ vertex $v$  is a face of a level $(n+1)$  vertex $w$, then the last vertex $u$ in $\Ver(w)$ is not present in $\Ver(v)$, but  is listed in $U_v$. From  Lines 12 and 21 of the algorithm it is clear that $U_w \subset U_v$ and moreover that $u \not \in U_v$. The cardinalities of the vectors $U_{(-)}$ are therefore decreasing for the newly created vertices. More precisely, for a vertex $t$ and its faces $s_i$, there exist $i$ such that $|U_t| \lneq |U_{s_i}|$. Level $n+1$ vertices are created only if there exist a level $n$ vertex $t$ such that $U_t \neq \emptyset$. Since the cardinality of the $U_{(-)}$  decreases with each iteration of the repeat loop, the algorithm will terminate. 

We remark finally that the size of the  directed flag complex corresponding to a given directed graph $\calg$ may be exponential in the size of $\calg$. In that case, the process of creation of a complex is usually stopped at some fixed dimension $n$. The time complexity of Algorithm~\ref{alg:generateDirectedFlagComplex} is proportional to the size of the output complex $\calh$, multiplied by maximal level of a vertex in $\calh$ (due to the target-source search performed in Line~\ref{alg:generateDirectedFlagComplex:targetSourceSearch}) of the algorithm.

\subsection*{ST2.4. Homology and Betti numbers.}
All homology computations carried out for this paper were made with $\mathbb{F}_2$ coefficients, using the boundary matrix reduced by an algorithm from the PHAT~\cite{phatURL} library.

\clearpage
\section{Supplementary Figures}

\newpage

\clearpage

\end{document}